\documentclass[journal=jctcce,manuscript=article,]{achemso}
\usepackage{mciteplus}

\input{utils/packages}

\newcolumntype{d}{D{.}{.}{1}}
\newcolumntype{L}{>{$}l<{$}}
\newcolumntype{R}{>{$}r<{$}}
\newcolumntype{C}{>{$}c<{$}}
\RequirePackage{dingbat}

\NewDocumentCommand\atu{m}{ 
    \SI{#1}{a.u.}\xspace%
}

\mathchardef\lt="313C \mathchardef\gt="313E
\mathcode`<="4268 \mathcode`>="5269
\newcolumntype{.}{D{.}{.}{-1}}
\newcolumntype{,}{D{,}{,}{2}}

\def\Put(#1,#2)#3{\leavevmode\makebox(0,0){\put(#1,#2){#3}}}

\def\hfsr{$E^\text{HF}_\text{SR}$\xspace}
\def\rmsanis{$\text{RMS}(e_\text{aniso})$\xspace}
\def\mseis{$\text{MSE}(e_\text{iso})$\xspace}

\def\chf{$c_\text{x}$\xspace}

\def\qqpolbase{Qpol}
\def\qqpol{Q\qqpolbase \xspace}
\def\qqpols{Q\qqpolbase s\xspace}

\def\Qqpols{Q\qqpolbase s\xspace}

\def\ddpolbase{Dpol}

\def\ddpols{D\ddpolbase s\xspace}

\def\mainbasis{da(wC)VQZ\xspace}
\def\psif{\texttt{Psi4}\xspace}

\def\errq{\Delta {Q}_{lm}} 
\def\errp{\Delta \alpha_{lmlm}} 
\def\eps{\epsilon}

\input{utils/utils}

\title[Pol-Benchmarks]
{
  Benchmarking higher-ranking multipoles and polarizability tensors for small molecular systems
}

\author{Bruno V. von Br\"uning}
\affiliation{Department of Physics and Astronomy, 
School of Physical and Chemical Sciences,
Queen Mary University of London,
London E1 4NS, U.K.}
\author{Alston J. Misquitta}
\affiliation{Department of Physics and Astronomy, 
School of Physical and Chemical Sciences,
Queen Mary University of London,
London E1 4NS, U.K.}

\email{a.j.misquitta@qmul.ac.uk}

\begin{document}

\date{\today}
\thispagestyle{empty}

\begin{abstract}
    Molecular properties govern how molecules interact with one another or external fields.
    Accurate molecular properties are essential for constructing intermolecular interaction models, and to achieve high accuracy we need to go beyond leading-order (dipolar) terms.
    
    This work presents CCSD(T) references for 73 small non-spin-polarized molecules for molecular dipole and quadrupole moments, and dipole--dipole and quadrupole--quadrupole polarizabilities.
    Using these results, we provide a holistic performance analysis of molecular properties computed with HF, MP2, CCSD, and a wide range of density-functional methods.
    Additionally, we investigate in detail what levels of basis sets are required to simultaneously describe all properties in the data set. 
    
    The best-performing density functionals for all four properties are the asymptotically corrected hybrid GGAs B97-3-AC, closely followed by PBE0-AC. 
    Surprisingly, modern meta-GGAs and range-separated functionals perform inconsistently, with large errors in the quadrupolar properties.
    These results have direct implications for methods for intermolecular interactions, such as symmetry-adapted perturbation theory based on density functional theory, or for the quality of properties and interactions in machine-learning datasets. 
    
    Finally, for DFT methods, the Jensen aug-pcseg-2 basis set is a computationally efficient alternative to more established basis sets, but for generating correlated references, third-row elements benefit from core polarization, and double augmentation is strictly necessary for quadrupole--quadrupole polarizabilities. 

\end{abstract}
\thispagestyle{empty}

\clearpage
\setcounter{page}{1}%

\section{
    Introduction
}\label{sec:intro}Molecular properties are key to both understanding and modeling intermolecular interactions. 
In fact, at long-range, the intermolecular interaction potential is fully determined by electrostatic multipoles and polarizability tensors.
\cite{Jeziorski1994,Stone2013}
Even at shorter-range interactions with significant density overlap, such models remain useful under appropriate damping and distribution to atomic sites.\cite{Stone2013,Misquitta2018}
This makes force field-based models parameterized through \textit{ab initio}-derived properties\cite{Bukowski2007,Totton2010,Misquitta2016a,Vandenbrande2016,VanVleet2016,VanVleet2018} a powerful alternative to empirical Lennard-Jones-like models.

The class of polarizable force fields, which model the response of molecules to external fields, has seen success in the field of biomolecular interactions.\cite{Jing2019,Melcr2019,Shi2015}
Some polarizable force fields, such as AMOEBA, have shown that \textit{ab initio}-derived electrostatic multipoles and polarization models can substantially improve accuracy.\cite{Holt2010,AkinOjo2013,Metz2016,Heindel2024,Gresh2007,Piquemal2006}
Moreover, one of the authors has demonstrated the systematic improvement achieved by including higher-ranking terms in the \textit{ab initio}-derived polarization expansion.\cite{Gilmore2019}
While classical force fields appear destined to be disrupted by machine-learned interatomic potentials, the consensus remains that long-range interactions can only be modeled through force-field-like terms.%
\cite{Esders2025,Kabylda2025b,Batatia2026,Parker2026}
This is commonly achieved by learning parameters for long-range interaction forms, namely electrostatics via partial charges and higher-order multipoles, and dispersion coefficients via atomic polarizabilities obtained through the Tkatchenko--Scheffler scheme.
\cite{Unke_2019,Thuerlemann2023a,Ple2025a,Moerman2026}

In principle, more accurate dispersion models may be obtained via \emph{ab initio}-derived dynamic polarizabilities, 
\cite{Misquitta2003,Misquitta2005,Shahbaz2019a} which could subsequently be machine-learned.\cite{Goennheimer2026a,Fang2025,Wilkins2019}
Indeed, this is what is done in ab initio methods like SAPT(DFT), which add dispersion through models using dynamic polarizabilities computed using linear-response time-dependent DFT,
\cite{Misquitta2005,Misquitta2005a,Hesselmann2004}
however many dispersion corrections for DFT methods still only feature the leading-order term.\cite{Grimme2006,Caldeweyher2019}
The leading $-C_6r^{-6}$ term arises from pairwise interaction and involves dynamic dipole-dipole polarizabilities (\ddpols). Higher-order terms, such as quadrupole-quadrupole polarizabilities (\qqpols), enter the $C_8$ and $C_{10}$ terms, which are known to account for around \SI{40}{\%} of the dispersion potential at equilibrium distances.\cite{Johnson2006,Misquitta2008a,Walters2018,Visscher2019,Misquitta2018} 
Similarly, for the leading electrostatic term, we must consider higher-order multipoles beyond the dipole. The question is: how do we compute these terms in a computationally efficient yet accurate manner across chemical space?

In practice, {\em ab initio}-derived multipoles and polarizabilities rely on Kohn-Sham DFT due to the large throughput needed.
Among the many existing Density Functional Approximations (DFAs), we may choose the one best suited to describe particular properties, but this is not acceptable if we wish for methods that work well across a range of properties. Additionally, thus far the focus has been only on leading-order properties.
For example, intermolecular potential datasets for machine learning, such as SPICE\cite{Eastman2023} and OMol25\cite{Levine2025}, employ the modern DFAs $\omega$B97M-V or $\omega$B97M-D3(BJ)\cite{Mardirossian2016} paired with def2-TZVPD and def2-TZVPPD basis sets.\cite{Weigend2005,Rappoport2010}
While modern DFAs can be trained to perform well on intermolecular benchmarks (see, for example, Ref.~\citenum{Mardirossian2016}), they do not necessarily improve the description of the density\cite{Sim2022, Medvedev2017, Brorsen2017} and polarizabilities\cite{Hait2018}.

Inaccuracies of properties produced by DFAs will certainly cause inaccuracies in intermolecular energies at long range;
but at short range they can also yield misleading energy-component analysis through schemes such as ALMO or SAPT(DFT)
(see, for example, \cite{Naseem_Khan_2022} for an analysis).
SAPT(DFT) constructs interaction energies explicitly from molecular properties, with higher-order terms being essential for accuracy.
The accuracy of SAPT(DFT) interaction energies therefore depends implicitly on using a DFA that provides a {\em holistic} description of molecular properties, meaning it faithfully describes both lower- and higher-ranking multipoles and polarizabilities.
Thus far, applications of SAPT(DFT) have relied largely on PBE0 combined with an asymptotic correction, but as we will demonstrate in this paper, the developments in DFAs over the last two decades have produced alternatives to PBE0.

To gauge a DFA's ability to yield a holistic representation of properties, we need to benchmark its performance against highly accurate references and cover a reasonably sized chemical space.
Previous studies --- which we will comment on more extensively later --- have provided references only for molecular dipoles and \ddpols, the leading-order terms for electrostatic interaction and induction/dispersion, respectively. In this work, we provide new references for traceless quadrupole moments and \qqpols, which, together with dipole moments and \ddpols, allow us to assess methods much more comprehensively.

The first comprehensive effort towards a high-accuracy benchmarking dataset of \ddpols was proposed by Hickey and Rowley at the CCSD/aug-cc-pVTZ level of theory.\cite{Hickey2014}
This was followed by a similar but higher-level study by Thakkar and coworkers employing the CCSD(T)/aug-cc-pVTZ level of theory.\cite{Wu2015}
Due to the high computational cost of CCSD(T), only the aug-cc-pVTZ basis could be employed for larger species.%
\cite{Joergensen2020,Hickey2014}
Hait and Head-Gordon presented the most accurate systematic studies at CCSD(T)/aug-cc-pCVQZ with basis set extrapolation, evaluating dipoles, \ddpols, and second cumulants.\cite{Hait2018,Hait2018a,Hait2021} 
The computational expense of this level of theory only permitted evaluation of species of up to 8 nuclei.

\Qqpols, on the other hand, have been computed at sufficiently high correlated levels only for a few molecules.\cite{Maroulis1990,Maroulis2003,Porsev2012,Kalugina2015,Loboda2016a,Kalugina2018}
Notably, experimental references for \qqpols have only been proposed for elemental ions.\cite{Sahoo2012,Higgins2021,Vylegzhanin2025} 
This is contrary to \ddpols, which have frequently been reported for atoms and molecules.\cite{Hohm2013,Johnson2002}
While computed atomic polarizabilities match their experimental counterparts,\cite{Cheng2024b}
comparing computed molecular polarizabilities to experiment requires accounting for vibrational effects.\cite{Monten2011,Huzak2013,Cassiano2025,Sharipov2017}
Furthermore, high-accuracy calculations at the CCSD(T) level of theory must be used, with perturbative triples contributing about \prc{2} to \ddpols.\cite{Hait2018}
Likewise, CCSD and CC3 \ddpols lie within  \SI{2}{\%} of experimental references for 14 monocyclic aromatic systems.\cite{Joergensen2020}
These discrepancies may be accounted for by vibrational effects observed in naphthalene and anthracene.\cite{Huzak2013}
Even for large conjugated systems like naphthalene and anthracene, CCSD(T) calculations match experimental results within \SI{1}{\%} when vibrational effects are accounted for, and basis set extrapolation is performed using a focal-point scheme.\cite{Huzak2013}
The effect of quadruple excitations has been explored only anecdotally for small basis sets, but it appears to be negligible.\cite{Hammond2009,Monten2011}
However, for specific compounds, Hait and Head-Gordon raise caveats on the sufficiency of CCSD(T) due to large deviations between CCSD and CCSD(T).\cite{Hait2018}
Finally, a non-relativistic Hamiltonian should suffice, with relativistic effects becoming significant only at the fourth row of the periodic table.%
\cite{Cheng2024b} 

In addition to the level of theory, basis-set completeness also matters.
For example, it is known that \ddpols are sensitive to augmentation by inclusion of diffuse functions; the so-called tail augmentation.\cite{Hickey2014} 
While the importance of single augmentation is widely acknowledged,\cite{Davidson1986,Hickey2014
} many studies also suggest the need for higher levels of augmentation.\cite{Woon1994,Osinga1997,Peterson1997,Elking2011}
It has been frequently suggested that the level of augmentation may have a higher impact than the basis set cardinality \cite{Monten2011,Huzak2013,Hurtado2024}
although the opposite has also been stated.\cite{Christiansen1998,BaranowskaLaczkowska2015,Joergensen2020}
Hait and Head-Gordon have also investigated the use of basis sets with core polarization, and proposed an improved description of core electrons via core polarization, but this was based on a study limited to the molecular dipoles of \ce{NH3} and \ce{PH3}.%
\cite{Hait2018,Puzzarini2008}
Likewise, a recent Hartree-Fock study suggests that core polarization significantly affects the \ddpols of second-row elements.\cite{Hurtado2024}
A study by Kalugina and coworkers found \qqpols of \ce{H2S} at the CCSD(T) level to be much more sensitive to the cardinality of aug-cc-pV$n$Z basis sets than \ddpols.\cite{Maroulis2003,Kalugina2018} 

In this paper, we provide static \qqpols and \ddpols, as well as dipole and quadrupoles at the CCSD(T) level, with a range of basis sets approaching the basis-set limit.
Since reference values for quadrupoles and \qqpols are novel, we first conduct a comprehensive analysis of basis-set convergence for a small set of molecules containing up to three atoms, using basis sets up to the pentuple-$\zeta$ level. 
In particular, we consider up to triply-augmented and core-polarized basis sets.
Having established the basis set requirements for each property, we propose a basis set appropriate for use as a reference across all four properties.

We then compute CCSD(T) references at this chosen basis set for a larger dataset containing 73 compounds.
For this data, we investigate how the properties vary across correlation levels, namely HF, MP2, and CCSD, before turning to the performance of a wide range of  density functional approximations, including the effect of an asymptotic correction for some of them. Throughout, our aim is a holistic analysis: in the DFA space, this means identifying the DFAs that perform well across all properties simultaneously, rather than excelling at some at the expense of others. We comment on trends in property accuracy along Jacob's ladder, examine the impact of the exact-exchange fraction, and finally assess how the choice of basis set affects DFA performance.

\section{
    Theory
}\label{sec:models}\subsection{Multipoles and Polarizabilities}
The interactions between a molecule and a time-dependent external field $V(\omega)$ with frequency $\omega$ can be described through perturbation theory. The first-order response entails electrostatic interactions that describe the interaction of the unperturbed electron density $\rho$ with the time-independent field $V=V(\omega=0)$:\cite{Parr1978}
\begin{align}
    \rho( \mathbf{r}) &= \frac{\partial E(V(\mathbf{r})=0)}{\partial V(\mathbf{r})}
\end{align}
Polarizabilities describe the second-order relaxation of the energy and can be computed using the charge density susceptibilities $\chi$:\cite[p30]{Stone2013}
\begin{align}
    \chi(\omega;\mathbf{r},\mathbf{r'})= \frac{\partial^2 E(V(\omega))}{\partial V(\omega;\mathbf{r}) \partial V(\omega;\mathbf{r'})}.
\end{align}
Multipolar electrostatic moments $Q_{lm}$ and polarizabilities $\alpha_{lml'm'}$ arise through a multipolar expansion of $V$ into moments $V_{lm}=v_{lm}M_{lm}$, with $M_{lm}$ being multipolar operators of rank $l$ and directional component $m$, and field strengths $v_{lm}$:
\begin{align}
Q_{lm} &=\frac{\partial E}{\partial V_{lm}}  \label{eq:def_mom}
\\ \alpha_{lml'm'} &= \frac{\partial^2 E}{\partial V_{lm} \partial V_{l'm'}}.  \label{eq:def_pol}
\end{align}

We only consider static polarizabilities ($\omega=0$) in this work, but previous studies on correlated and DFT-based methods suggest that errors in static polarizabilities are reasonably transferable to dynamic polarizabilities ($\omega \gt 0$).\cite{Joergensen2020,Grotjahn2020,Beizaei2021} 
Thus, the quality of static polarizabilities can serve as a reasonable proxy for dynamic polarizabilities.
However, the literature suggests much more limited transferability from polarizabilities to higher-order responses to external fields, termed hyperpolarizabilities.\cite{Salek2005,Karne2015,Zalesny2019}
The simplest way to obtain $Q_{lm}$ and $\alpha_{lml'm'}$ is through multiplication of $\rho$ and $\chi$ with the corresponding multipolar operators $M_{lm}$. Importantly, this approach also allows deriving distributed multipoles and polarizabilities via either basis- or real-space partitioning.

\subsection{Finite Field Calculations}
Perturbative methods, such as CCSD(T), do not yield orbital coefficients in their first-order expansion, which does not allow the evaluation of $\rho$ and $\chi$. Likewise, semi-local DFAs introduce inaccuracies through the adiabatic approximation in linear-response calculations and also depend on the implementation of the corresponding response kernels in the respective quantum chemical code.
In these cases, one can also obtain $Q_{lm}$ and $\alpha_{lml'm'}$ through finite difference calculations with respect to equations \eqref{eq:def_mom} and \eqref{eq:def_pol}:
\setlength{\arraycolsep}{0pt}
\begin{alignat}{4}
    Q_{lm}  &= &&\ \hspace{.75cm} \frac{E( fM_{lm}) - E(-f M_{lm})}{2f}  &+\ &\mathcal{O}(f^2)  \label{eq:ff_def_mul}
    \\ \alpha_{lmlm} &= && \ \frac{E(f M_{lm}) - 2 E(0) + E(- f M_{lm})}{f^2} \ &+\  &\mathcal{O}(f^2) \label{eq:ff_def_pol}
\end{alignat}
where $E(fM_{lm})$ is the relaxed energy in the external field $fM_{lm}$, with $f$ the finite field strength.
As indicated in equation \eqref{eq:ff_def_pol},
we only consider the diagonal elements of the polarizabilities, but in principle one could obtain arbitrary elements $\alpha_{lml'm'}$ as outlined in \ref{si:ff_setup}. 
The multipolar operators considered in this work are the \emph{Cartesian} dipole operator $M_a$ and the traceless quadrupole operator $M_{ab}$:
\begin{align}
    M_a &=r_a  \label{eq:dip_field}
    \\ M_{ab} &= \left(r_ar_b - \frac{1}{3}\delta_{ab}r^2\right) 
    \label{eq:qad_field}
\end{align}
where $r_a,r_b$ denote principal Cartesian components and $\delta_{ab}$ the \emph{ Kronecker} Delta.
Notably, quadrupolar properties depend on the origin of the field expansion, which we define as the center of nuclear charge. 

For non-perturbative methods -- like semi-local or hybrid DFAs -- one can obtain polarizabilities through a finite difference by 
 evaluating relaxed multipole moments $Q_{lm}$ within a field $V_{l'm'}$.\cite{Elking2011}
\begin{align}
    \alpha_{lml'm'}= \frac{Q_{lm}(f M_{l'm'}) - Q_{lm}(-f M_{l'm'})}{f} + \mathcal{O}(f^2) \label{eq:ff_pol_from_dens}
\end{align}
This would yield all desired off-diagonal elements $lm,l'm'$  for application of a given field $fM_{l'm'}$, including dipole-quadrupole and dipole-octopole polarizabilities.

\subsection{Choice of Finite Field Value}

Numerical errors of finite-field calculations on multipoles $\errq$ and polarizabilities $\errp$ can be controlled by the choice of $f$. 
Large $f$ pollute multipoles through first-order hyperpolarizabilities $\beta$, and polarizabilities through second-order hyperpolarizabilities $\gamma$ (see \ref{si:ff_errors}):
\begin{align}
    \errq &=\frac{\beta f^2}{2} + \mathcal{O}(f^3) 
    \\\errp &=\frac{\gamma f^2}{2} + \mathcal{O}(f^3)
\end{align}
One should furthermore not choose $f$ so large that the external field enables unbound electronic states.

Small $f$ leads to finite field changes in the energy that are on the order of the numerical uncertainties of the SCF energy, which are bounded by the convergence criteria.
Assuming that $\eps$ is the worst-case magnitude of the uncertainty, we can estimate for small $f$: 
\begin{align}
    \left(\errq\right)_{f\rightarrow 0} &\le \frac{2\eps}{f} \label{eq:ff_uncertainty_mul}
    \\ \left(\errp\right)_{f\rightarrow 0} &\le \frac{4\eps}{f^2}\label{eq:ff_uncertainty_pol}
\end{align}
Conversely, we can choose finite field values  for multipoles $f_Q$ and polarizabilities $f_\alpha$ for given tolerances $\Delta Q^\text{tol}$ and $\Delta \alpha^\text{tol}$, and $\epsilon$:
\begin{align}
    f_Q\le\frac{2\eps}{\Delta Q^\text{tol}} 
    \\ f_\alpha\le \frac{2\sqrt{\eps}}{\Delta \alpha^\text{tol}}
\end{align}
Since $ \eps \ll 1$, for a given $\eps$ the admissible $f_Q$ is much larger than $f_\alpha$.
We will use the inequalities below to determine the finite-field settings.

\subsection{Origin of Properties and Asymptotic Correction}
This work investigates the impact of correlation treatment and basis-set flexibility on properties, so it is fair to ask which electronic-structure effects govern multipoles and polarizabilities.
While we will provide a much more detailed analysis of polarizabilities below, we can only say that the picture for multipoles may be somewhat ambivalent due to possibly competing effects of interatomic charge transfer (effective partial charges), anisotropic polarization of electronic density around atomic centers, and the nature of bonding orbitals.
Previous authors suggested that increasing occupied-unoccupied energy gaps resulting from correlation could increase antibonding character and hence lower dipoles.\cite{Hait2018a}

The $r$-dependence of multipolar operators differently amplifies the sensitivity of properties to regions far from the center of expansion:
$r^1$ for dipoles, $r^2$ for quadrupoles and \ddpols, and $r^4$ for \qqpols.
Thus, we expect the quadrupoles and polarizabilities, particularly the \qqpols, to be sensitive to the density tails and, consequently, to correlation effects associated with the valence states.
Polarizabilities additionally depend on the excitation spectrum, particularly the HOMO--LUMO gap, as virtual states enter the charge-density susceptibility functions.

(Semi-)local Kohn-Sham DFT overestimates the density diffuseness because the one-particle potential has the wrong asymptotic form.\cite{Tozer1998,Casida1998,Gruening2001, Salek2005,Cencek2013,Wu2015}
For a neutrally charged atomic system, electrons should be attracted to the nucleus by a $-1/r$ potential at long range, but in a self-interacting potential the electrostatic potential is screened and tends to $0$ exponentially with separation.\cite{Tozer1998,Casida1998,Gruening2001}
This self-interaction error can be corrected by imposing the asymptotically correct long-range form of the one-electron potential $V_{r\rightarrow\infty}$, which includes a shift arising from the derivative discontinuity in the energy at integer numbers of electrons \cite{Perdew1982}:
\begin{align}
    V_{r\rightarrow\infty} = -\frac{1}{r} + 
    I + E_{\text{HOMO}},
\end{align}
where $I$ is the first vertical ionization energy, and $E_{\text{HOMO}}$ is the highest occupied molecular orbital energy. 

A popular scheme to enforce the correct $V_{r\rightarrow\infty}$ is the gradient-regulated asymptotic connection (GRAC)\cite{Gruening2001}, which interpolates between bulk and low-density regions through a two-fixed-parameter model based on the density and reduced density gradient.
GRAC is important for SAPT(DFT),\cite{Hesselmann2002,Misquitta2005} and has been shown to improve \ddpols.\cite{Gruening2001,Gisbergen1998,Bast2008,Cencek2013}
$E_{\text{HOMO}}$ and $I$ appear as compound-specific parameters in GRAC, and in principle, these can be determined self-consistently; however, we use a non-self-consistent approach, using uncorrected DFA to determine $I$ from the DFT energies of the ionized and neutral compounds, and $E_\text{HOMO}$ from the ground state.

\section{
   Methods
}\label{sec:numerical}\subsection{Chosen Compounds}
We computed properties only for neutral systems in this work, but note that properties of charged compounds are certainly important for interaction models. For anions, this will likely require particular care in basis set convergence studies due to the inherent diffuseness of their density.
Furthermore, we did not consider elements beyond the third row since they can exhibit significant relativistic effects.\cite{Cheng2024b}
Lastly, we considered only non-spin-polarized species that allow treatment with a restricted HF determinant. 

For our main dataset, we considered 73 compounds with up to 7 nuclei and an average of \SI{3.8}{} nuclei, listed in Table \ref{tab-si:compound_for_main_dataset}. The four atoms \ce{H}, \ce{He}, \ce{Ne}, and \ce{Ar} were considered and of the 69 remaining compounds,
45 contained either \ce{N}, \ce{O}, or \ce{F}, 29 compounds contained carbon, 28 compounds contained either \ce{Si}, \ce{P}, \ce{S}, or \ce{Cl}, and 13 compounds contained \ce{B}. No other elements appeared in the data.
We sourced all these compounds from Ref.~\citenum{Hait2018} and Ref.~\citenum{Hait2018a}, where they were identified as non-spin-polarized through stability analysis.
Notably, many of the spin-polarized species identified in Ref.~\citenum{Hait2018} are challenging, as shown by large DFA errors in \ddpols.
We note that we discarded some non-spin-polarized species found in Ref.~\citenum{Hait2018} and Ref.~\citenum{Hait2018a}, including molecular complexes, metal-organic compounds, and ionically bound species (see \ref{si:dataset-compounds}).  

The high cost of our basis-set convergence study at the CCSD(T) level limited us to 22 species, whose identities are shown in the legend of Figure \ref{fig:basis_set_convergence}. Of these 22 species, 18 come from the previously described non-spin-polarized subset.
We added \ce{F2}, \ce{P2}, and \ce{PN} from the spin-polarized subset found in Ref.~\citenum{Hait2018}, but described them with a restricted reference. We also included \ce{SiS}, which is not found in Ref.~\citenum{Hait2018}, Ref.~\citenum{Hait2018a}, nor in Ref.~\citenum{Hait2021}, using an experimental structure from the NIST database.\cite{Johnson2002} 
\ce{SiS} is isoelectronic to \ce{P2}, and like \ce{P2}, it is expected to fall in the spin-polarized category.

\subsection{Chosen DFAs}
We used all DFAs found in \psif, but note that this does not represent all common DFAs, with notable examples being mBEEF and $\omega$M05-D6, which performed well on \ddpols in reference \citenum{Hait2018}.
We considered only base DFAs, not different dispersion-corrected implementations, except for $\omega$B97X\cite{Mardirossian2014}, B97M\cite{Mardirossian2015}, and $\omega$B97M\cite{Mardirossian2016}, which were consistently parametrized with VV10 and D3(BJ). They generally perform similarly, and this work refers only to the -V version.

In total, 176 DFAs have been considered in this work; their count by class is listed in Table \ref{tab-si:dfa_count} and their identities in Table \ref{tab-si:dfa_identities}. This includes 14 DFAs for which we applied the GRAC correction, which we denote with the suffix -AC; these are identified by the red marking of their names in Table \ref{tab:methods_leaderboard}. Including the GRAC-corrected version. The total count of 176 DFAs does not include 17 DFAs we discarded due to convergence issues, predominantly derivatives of the mGGA B95 correlation kernel. This is unfortunate since the PW6B95 mGGA previously showed great performance for dipoles and \ddpols in Ref.~\cite{Hait2018a} and Ref.~\cite{Hait2018}.
For the remaining DFAs, convergence problems occurred rather infrequently. 
Further details on sampled DFAs and convergence failures can be found in \ref{si:dfas}.

\subsection{Calculation settings}

All calculations were performed with the open-source package \psif (version 1.10), using default settings unless otherwise noted.
We describe energy convergence thresholds in the finite-field setup.
DFT grid settings match Ref.~\citenum{Hait2018} and Ref.~\citenum{Hait2018a}: 99 radial grid points in the Treutler scheme and 590 angular grid points in the Lebedev scheme.
Grids for the VV10 correction followed default \psif parameters.

We used various versions of \textit{Dunning}-type basis sets throughout this work,\cite{Dunning1989}
including the standard singly-augmented aug-cc-pV$n$Z as well as the doubly- and triply-augmented
d-aug-cc-pV$n$Z and t-aug-cc-pV$n$Z variants,\cite{Woon1994} which we denote as aV$n$Z, daV$n$Z, and taV$n$Z, respectively. These higher-augmented sets, along with
higher-augmented \textit{Jensen} and core-polarized \textit{Dunning} basis sets, were generated using the \texttt{Python} package \texttt{Basis Set Exchange} (version 0.11).\cite{Pritchard2019}
We also used the core-polarized aug-cc-pCV$n$Z and weighted core-polarized aug-cc-pwCV$n$Z variants of the aV$n$Z basis\cite{Woon1995,Peterson2002},
which we denote as aCV$n$Z and awCV$n$Z in the following, the latter employing slightly more diffuse functions to describe core flexibility.

Because conventional CCSD(T) requires large amounts of memory and disk space, we used density fitting for the main dataset.\cite{Baerends1973,Whitten1973} 
For the Dunning-type basis sets, we employed the corresponding resolution-of-identity density-fitting basis sets developed by Weigend and coworkers.\cite{Weigend2002,Haettig2005} Since doubly-augmented versions of these fitting basis sets are not available, we used the corresponding singly-augmented version for each basis set. The density-fitting basis sets for \textit{non-Dunning}-type basis sets are described in \ref{si:df-basis_sets}.
The deviations between the density-fitted -- with the aforementioned density-fitting basis -- and conventional CCSD(T) were assessed for the compounds of the basis set convergence study: the deviations lie strictly below \prc{0.25} (see also Figure \ref{fig-si:density-fitting-error}). Previous studies have shown that the relative density-fitting error remains constant with system size \cite{Werner2003,Sodt2006,Hollman2014}, and we expect comparable density-fitting errors for computation of our main dataset.

We note that Cholesky decomposition\cite{Aquilante2007,DePrince2013} and frozen natural-orbital approximations\cite{Crawford2019} would improve computational efficiency, but we could not use these approximations in our finite-field calculation because their active spaces jump when applying external fields. Likewise, the frozen-core approximation can cause notable deviations, in particular for third-row atoms (see discussion in \ref{si:df}).

\subsection{Finite Field Setup}
The dipolar and traceless quadrupolar fields (see equations \eqref{eq:dip_field} and \eqref{eq:qad_field}) were implemented through an arrangement of point charges, which is specified in \ref{si-ff-setup}. Finite field values of \atu{1.e-4} were chosen for all calculations. 
For all compounds with up to three nuclei, convergence thresholds of \atu{1.e-10} were applied to energy difference (\texttt{e\_convergence}), RMS difference of CC solution vector (\texttt{r\_convergence}), and orbital gradient (\texttt{d\_convergence}). For all compounds with four or more nuclei, these thresholds were set to \atu{1.e-09}.
As suggested by equation \eqref{eq:ff_uncertainty_pol}, this should limit errors to below \atu{.01} for energy convergence of \atu{1e-10} and \atu{.1} for energy convergence of \atu{1.e-9}. 

We validated the finite field setup for polarizabilities by comparing finite-field HF calculations with linear response calculations: disagreements do not exceed \SI{.005}{\%} (see also figure \ref{fig-si:ff-vs-lr_main}).
Notably, this indicates a much tighter convergence than previously suggested by the energy threshold, which might be due to stricter convergence imposed by the convergence threshold on orbitals (\texttt{d\_convergence} in \psif).
Furthermore, we compared finite-field multipoles against multipoles evaluated through the density at HF level; disagreements do not exceed \SI{0.015}{\%}. While still negligibly small, the larger disagreement for multipoles is surprising and may arise from \psif integration thresholds.
Because \psif does not apply GRAC to energies, we evaluate finite-field calculations for DFA at the density level according to equation \eqref{eq:ff_pol_from_dens}.

\section{
    Analysis
}\label{sec:analysis}\subsection{Error Metrics}
Polarizabilities and multipoles are tensorial properties, and each component can imply chemical anisotropy. However, unlike the analysis of individual systems, our ensemble analysis does not permit inspection of individual components; instead, it requires condensing tensorial properties into scalar indicators.
Generally, one defines an isotropic norm -- such as the magnitude of a dipole or the trace of a higher-order multipole or polarizability tensor -- together with a supporting anisotropic indicator. For tensors, the latter could be the Frobenius norm \cite[p.27]{Stone2013}, which provides an upper bound on the eigenvalues without requiring their explicit computation. For polarizabilities, however, we did not evaluate the full tensor in order to reduce computational cost, and hence cannot determine either the Frobenius norm or the eigenvalues; consequently, we define the following error metrics using only the diagonal terms.

First, we define the magnitude of multipoles and polarizability tensors using isotropic norms:
\begin{align}
    \text{mag}(\theta) &=
        \left\lbrace\begin{array}{cl}
         \sqrt{\sum_\gamma (\theta_{\gamma})^2}&\quad \text{if multipole} \\
         \sum_\gamma |\theta_\gamma| &\quad \text{if polarizability}
        \end{array}\right.
\end{align}
Due to their chemical nature, polarizability elements are generally positive (but their differences are considered in the following); hence, the definition of their magnitude above is equal to the trace.
Using these magnitudes, we define the relative isotropic error $e_\text{aniso}$ between a trial tensor $\theta$ and a reference tensor $\theta'$:
\begin{align}
    e_\text{iso}(\theta,\theta') &= \frac{\text{mag}(\theta)-\text{mag}(\theta')}{\text{mag}(\theta')}.
\end{align}
We define the anisotropic error metric as:
\begin{align} \label{eq:def_diagonal_error}
    e_\text{aniso}(\theta,\theta')&= \frac{\text{mag} (\theta-\theta')}{\text{mag}(\theta')} \cdot \text{sign}(e_\text{iso})
    \left\lbrace\begin{array}{cl}
    =\frac{\sqrt{\sum_\gamma ((\theta-\theta')_\gamma)^2 }}{\sqrt{\sum_\gamma (\theta'_\gamma)^2}}\cdot \text{sign}(e_\text{iso})
    & \quad \text{if dipole or quadrupole}\\
    =\frac{ \sum_\gamma |\theta_{\gamma}-\theta'_{\gamma}| } 
    {\sum_\gamma  |\theta_{\gamma}'|} 
    \cdot \text{sign}\Big(e_\text{iso}\Big). &\quad \text{if polarizability} \\
    \end{array}\right.
\end{align}
Note that the error metrics for dipoles and quadrupoles are rotationally invariant, while polarizabilities depend on the choice of coordinates.

\subsection{Basis Set Convergence}\label{sec:cc_basis_set_convergence}
\begin{figure}
    \centering
    \includegraphics[width=\linewidth]{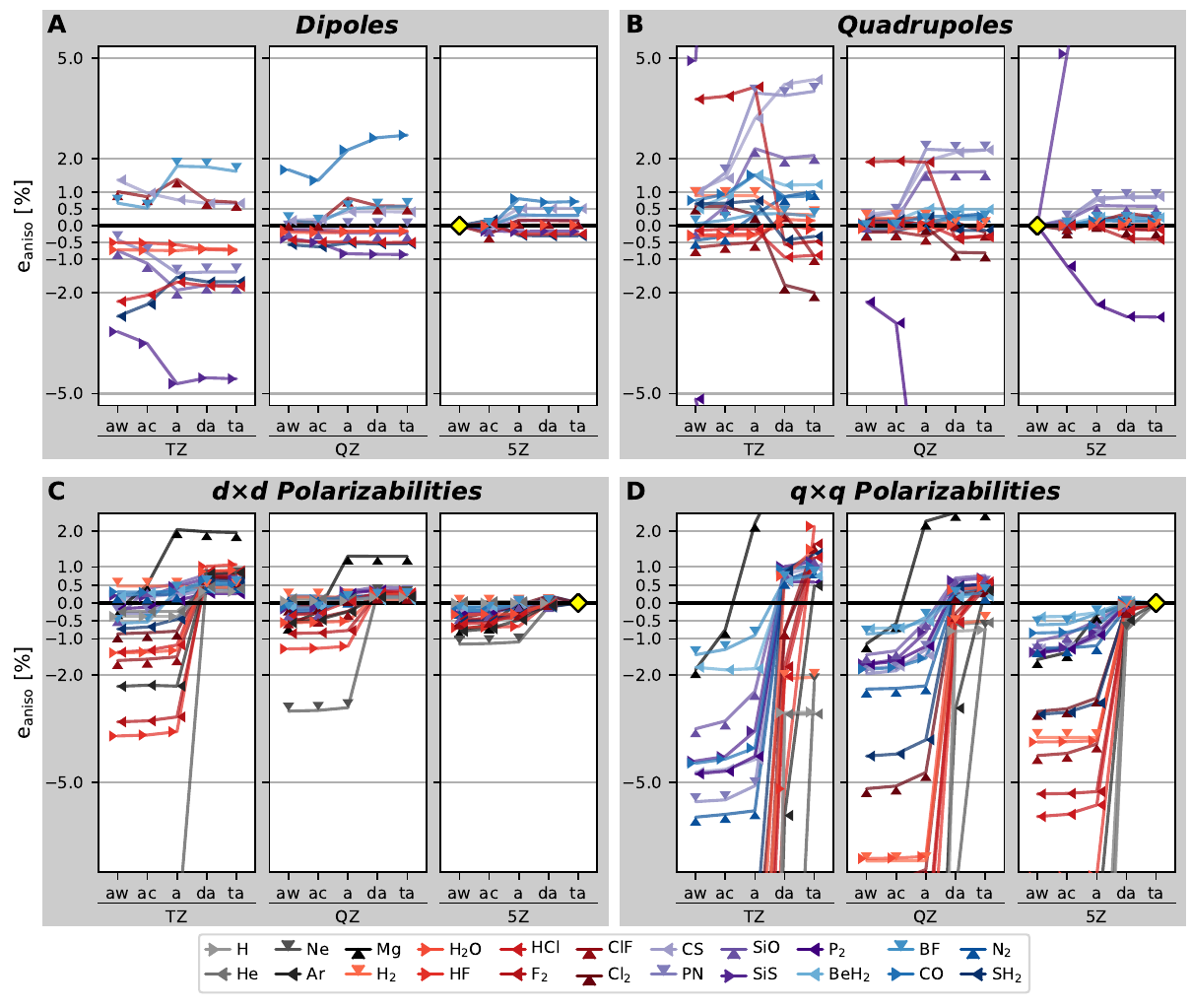}
    \caption{Anisotropic errors compared to the largest considered basis set. This is t-aug-cc-pV5Z for polarizabilities and aug-cc-pwcV5Z for multipoles. Single, double, and triple augmentation are indicated by a, da, ta, while wc and c indicate weighted and standard core polarization.
    \label{fig:basis_set_convergence}
    }
\end{figure}
\begin{figure}
    \centering
    \includegraphics[width=0.75\linewidth]{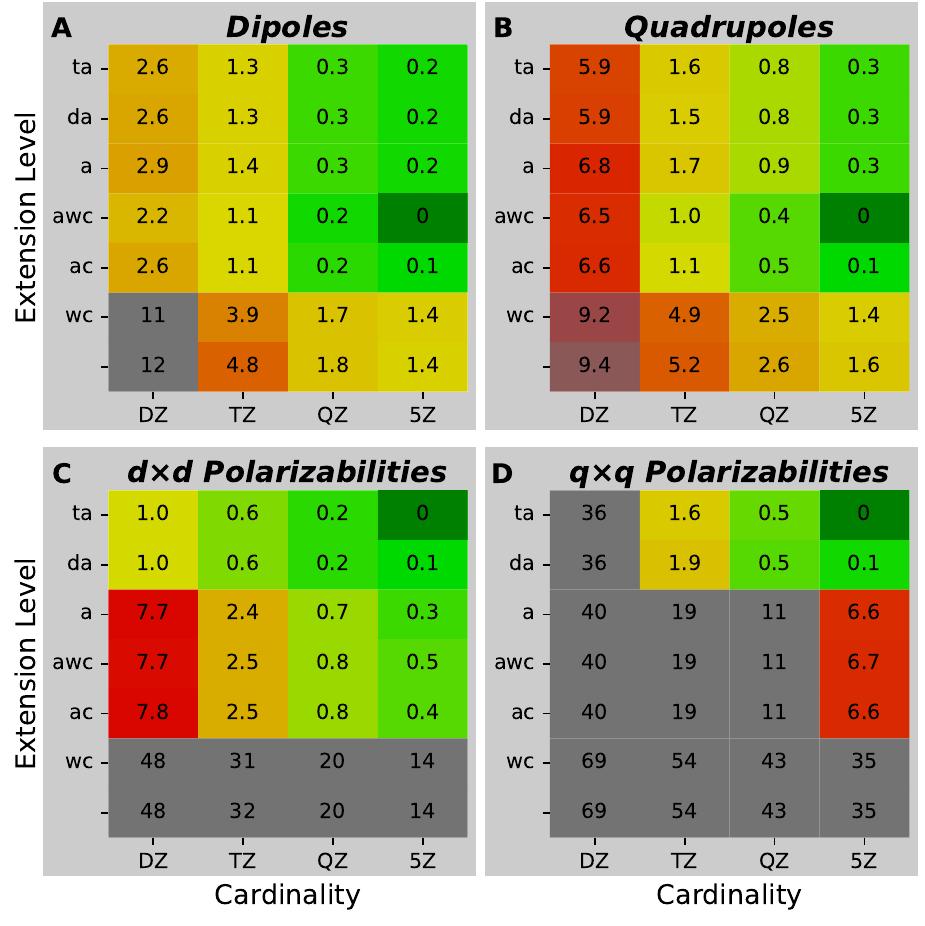}
    \caption{
    RMS of relative anisotropic error against the reference basis (marked deep green) at CCSD(T).
    The particles shown in the extension level denote single, double, and triple augmentation by a, da, and ta, while weighted and conventional core polarization are denoted at wc and c. 
    The following outliers were removed from the averaging: \ce{CO} for dipoles, \ce{SiS} and \ce{P2} for quadrupoles, \ce{Mg} for \ddpols, and \ce{Ne}, \ce{Ar}, \ce{Mg} for \qqpols.
    }
    \label{fig:heatmap_basis_set}
\end{figure}

To assess basis-set quality, we examine the convergence of each property with increasingly large basis sets, paying particular attention to the role of augmentation and core polarization.
The largest basis sets considered are taV5Z, which features a highly augmented tail, and awCV5Z and aCV5Z, which show the highest augmented core.
A priori, it is not clear which of these choices --- tail-augmentation or core-polarization --- will be more appropriate for each of the four properties we compute. 
Since combining triple augmentation and core polarization at the pentuple-$\zeta$ level is too computationally costly, we instead investigate them separately.
As outlined in the introduction, evidence suggests that tail-augmentation is important for polarizabilities, but no systematic tests have examined the role of core polarization, although previous studies have raised this question \cite{Hait2018,Hait2018a,Hurtado2024}.

In Figure \ref{fig:basis_set_convergence} we show $e_\text{aniso}$ versus the reference basis  --- marked by the yellow diamond --- for a selection of smaller basis sets.
We will explain how we determine the reference basis shortly.
For each cardinality $n$, we present five kinds of basis extension: the regular aV$n$Z basis set is in the center; on the left are the core-polarized variants, and on the right are the tail-augmented variants.
To determine the reference basis sets, we now consider only the pentuple-$\zeta$ data.

For dipoles (Figure \ref{fig:basis_set_convergence}A), going from the aV5Z level to the d-aug- or t-aug variant causes almost no change.
This indicates that the dipoles are almost unaffected by tail augmentation (at this basis cardinality). 
In contrast, going from aV5Z to the core-polarized variants of this basis causes a small but noticeable change. This shows that core polarization affects dipoles, as suggested by previous works.\cite{Halkier1999,Puzzarini2008}

For quadrupoles (Figure \ref{fig:basis_set_convergence}B), we see a much stronger effect of core polarization on $e_\text{aniso}$, while the effect of higher levels of tail-augmentation is negligible.
Some molecules, such as \ce{P2} and \ce{SiS}, show marked changes in their quadrupoles upon inclusion of core polarization. 
Apart from these two cases, both weighted core polarized (pwCV$n$Z) and conventional core polarization (pCV$n$Z) basis sets perform similarly at the quadruple-$\zeta$ and pentuple-$\zeta$ levels and therefore provide an equally good account of core polarization. Their similar performance also suggests good convergence with respect to core polarization.
We recommend weighted core polarization since, in most instances, it yields larger changes in errors than the corresponding standard aV$n$Z level, which may indicate a more complete account of core effects.
On this basis, we argue that the core-polarized basis awCV5Z should be the reference for the dipole and quadrupoles.

Figure \ref{fig:basis_set_convergence} shows the data for \ddpols and \qqpols in panels C and D. Once again, we focus on the pentuple-$\zeta$ basis sets. 
This time we see that with respect to the standard aV5Z basis, the core-polarized basis sets show almost no change in error, while going to the d-aug- and t-aug- basis sets has a much larger effect; in particular, even at pentuple-$\zeta$ level, double augmentation can change \qqpols by more than \prc{5} for many of the considered compounds. Notably, the small change between the double- and triple-augmented levels confirms convergence of polarizabilities at the triple-augmented level. 
Thus, for the polarizability tensors, we use the taV5Z basis data as reference values.

The sensitivity of properties to the basis set varies among compounds, with notable groupings of compounds indicated by coloring in Figure \ref{fig:basis_set_convergence}. First, we note that the larger fluctuations in the dipole error for \ce{CO} can be considered artifacts arising from its notoriously small dipole magnitude of around \atu{0.1}.
Multipoles of compounds containing the third-row elements, \ce{Si}, \ce{P}, \ce{S}, which are color-coded violet, show increased sensitivity to core polarization. While the sensitivity of these systems for dipoles is smaller -- but more noticeable at quadruple-$\zeta$-level --,
quadrupoles of these systems are quite sensitive even at pentuple-$\zeta$-level, particularly \ce{SiS} and \ce{P2}.
While dipoles are virtually insensitive to double augmentation, the quadrupoles of \ce{F2} and \ce{Cl2} change notably.
Notably, the large sensitivity of \ce{SiS} and \ce{P2} to core polarization, which already occurs at the HF level, may be due to their expected multireference character, which may be sensitive to core polarization. 
Nevertheless, this might warrant revisiting their \textit{Dunning} basis sets for third-row elements.

The polarizabilities of many species containing halogens or hydrogens (highlighted in shades of red and grey) are very sensitive to tail augmentation.
Noble gas atoms and \ce{H} are outstandingly sensitive to higher levels of tail augmentation.
As with multipoles, polarizabilities of compounds containing the third-row elements \ce{Si}, \ce{P}, and \ce{S} show sensitivity to core polarization, albeit by much smaller margins. 
Furthermore, the polarizabilities of \ce{Mg} require core polarization, which can be explained by the proximity of the 3s shell to the core electrons, as suggested by previous authors.\cite{Hait2018} 
As previous authors have observed \cite{Hurtado2024}, doubly and triply augmented basis sets with smaller $\zeta$ values overestimate polarizabilities, as is particularly evident at the daVDZ level for \ddpols in figure \ref{fig:basis_set_convergence}A. 
We generally observe that polarizabilities increase strongly with tail flexibility and decrease slightly with core flexibility.

\subsubsection{Recommended Basis Set for Correlated Calculations}

Figure \ref{fig:heatmap_basis_set} summarises the results from the previous section, reporting relative \rmsanis computed with respect to the reference basis set for each of the four properties.
As explained above, for multipoles the reference basis is awCV5Z, and for the polarizabilities it is taV5Z. 
From this figure, we see that the smallest acceptable basis for multipoles is awCVQZ (relative RMS errors less than \prc{0.5}), while for polarizabilities we need daVQZ (relative RMS errors of \prc{0.2} and \prc{0.5}). 
Since the need for core-augmentation arises primarily from compounds containing the third-row elements \ce{Si}, \ce{P}, and \ce{S}, we propose a composite basis that uses the daVQZ basis sets for all but the third-row atoms, for which the corresponding core-augmented basis, dawCVQZ, is used. In retrospect, we would also recommend using core polarization at \ce{Cl} nuclei.
We refer to this composite basis as \mainbasis, and we expect $e_\text{aniso}$ due to basis-set incompleteness at the \mainbasis level to remain below \prc{0.5} for all properties; we will use this basis in subsequent benchmarks on the larger dataset.

\subsection{Performance by Method}

We will now analyze the choice of method at the \mainbasis level.
Before analyzing DFAs, we first examine the impact of properties through different levels of correlation; namely, HF, MP2, and CCSD. We note that while polarizabilities of \ce{H} have been computed, there are substantial outliers for DFA -- as previously observed in Ref.~\citenum{Hait2018} -- and hence are omitted from the calculation of \rmsanis and \mseis for consistency. Furthermore, we omit the dipole of \ce{CO} and the quadrupole of \ce{BHO} since their unusually small magnitudes cause artifacts in the relative errors.

\subsubsection{Impact of correlation}
\begin{table*}
    \centering
    \includegraphics[width=.475\textwidth,valign=t]{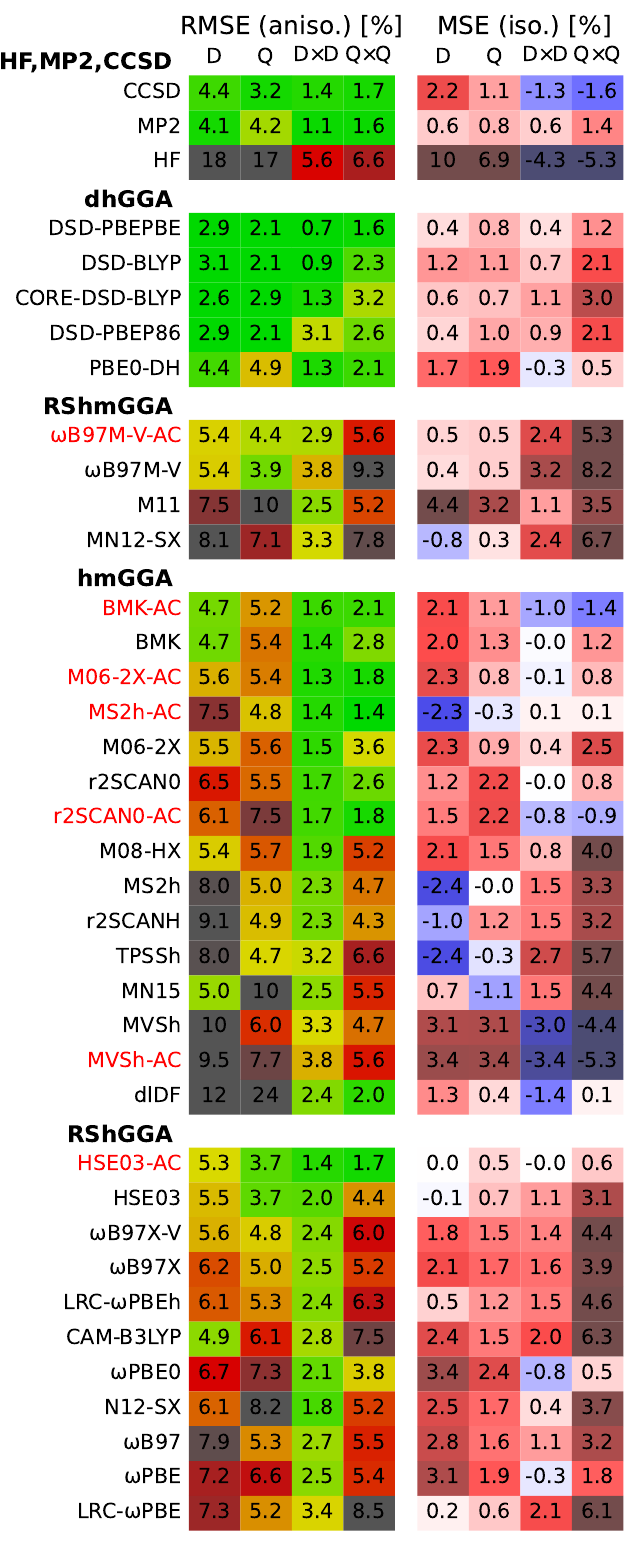}\hfill
    \includegraphics[width=.475\textwidth,valign=t]{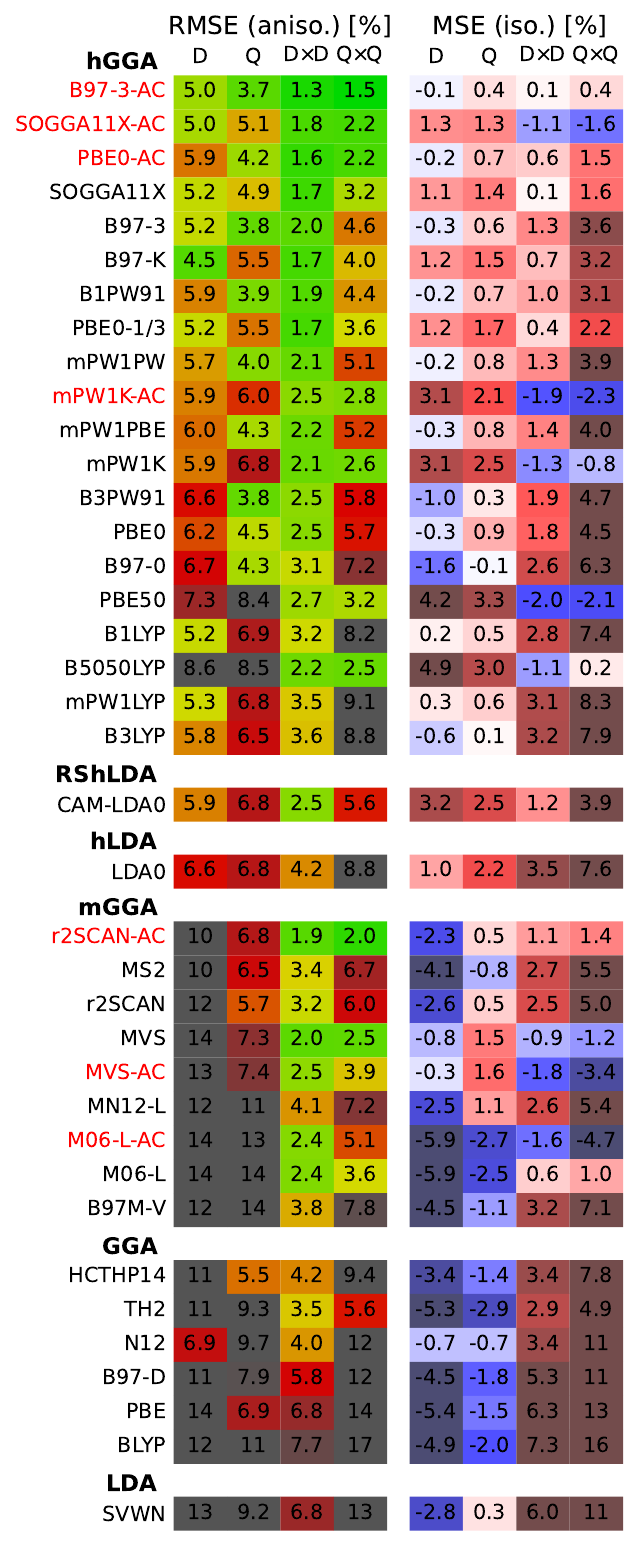}
    \caption{
    \rmsanis and \mseis 
    for  dipoles, quadrupoles, \ddpols, and \qqpols, which are denoted as D, Q, D$\times$D, Q$\times$Q, respectively. The color scheme differs for each property and stretches from \prc{4} (D), \prc{3} (Q), \prc{1} (D$\times$D), \prc{1.5} (Q$\times$Q) to \prc{8}. Within each class, DFAs are sorted by their aggregate performance aligned with this color scheme.
    The authors selected the DFAs presented in this subsection at their discretion; a full list appears in \ref{tab-si:methods_leaderboard}.
    \label{tab:methods_leaderboard}
    }
\end{table*}
\begin{figure}
    \centering
    \includegraphics[width=\linewidth]{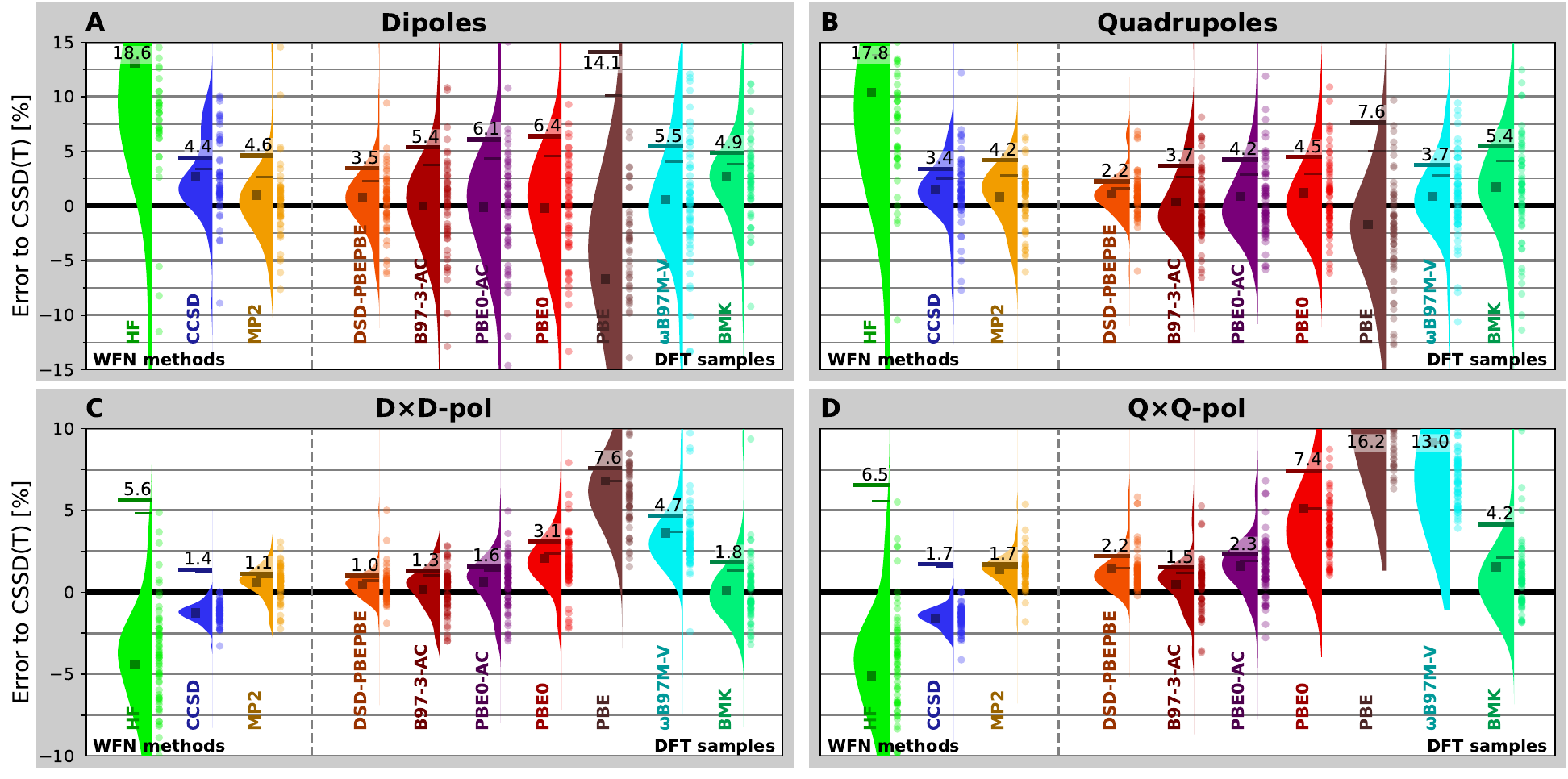}
    \caption{Distribution of $e_\text{aniso}$ made against CCSD(T) at \mainbasis. 
    For each distribution, \rmsanis is specified by a number and its position marked by a long, thick horizontal dash; MAE($e_\text{aniso}$) is located with a shorter, thinner dash; and $\text{MSE}(e_\text{aniso})$ is indicated by the square. 
    }
    \label{fig:error_distribution}
\end{figure}

Figure \ref{fig:error_distribution} shows the distribution of $e_\text{aniso}$ and \rmsanis for each method and property. 
First, \rmsanis is about three times larger on multipoles than on polarizabilities. 
Thus, correlation impacts multipoles much more strongly than polarizabilities. 
For all properties, MP2 and CCSD perform similarly, and both lower \rmsanis by about \prc{75} compared to HF. 
This indicates that double excitations account for \prc{75} of the correlation effect, and triple excitations account for the remaining \prc{25}. This \prc{75} to \prc{25} relationship holds remarkably well even for individual compounds (see figure \ref{fig-si:correl_part_by_compound}).
The convergence of properties with correlation appears hence quite systematic. The sensitivity of properties to correlation, on the other hand, varies widely between compounds, as seen by the wide error distribution of HF in figure \ref{fig:error_distribution}.

HF systematically underestimates polarizabilities and overestimates multipoles; CCSD strictly underestimates; and MP2 overestimates polarizabilities, with a more mixed picture for multipoles. In general, correlation increases polarizabilities; however, MP2 tends to overestimate polarizabilities, as has been observed (indirectly) by previous studies on the dispersion energy \cite{Pitonak2009,Nguyen2020}.
For multipoles, double excitations appear to strongly lower multipoles, while triple excitations appear to have a more mixed effect. 
Hence, the impact of correlation on multipoles appears more subtle than for polarizabilities.
Notable exceptions include the overestimation of HF-level \ddpols for \ce{C2H4}, \ce{C2H2}, and \ce{CH2BH}. These species are distinguished by their double bond, suggesting that non-local correlation due to delocalized electrons may be the cause.

The error distributions for both polarizabilities and multipoles are relatively tightly clustered, and while outliers exist, they do not dominate the distributions. \rmsanis is therefore a good indicator of the sampled ensemble error, and it exceeds the MAE by no more than about \prc{50}. 
While MP2 and CCSD errors are well clustered, particularly for the polarizabilities, CCSD dipoles and, to a lesser extent, quadrupoles show a bimodal distribution, as seen in panels A and B of Figure \ref{fig:error_distribution}. We have not been able to associate this with an obvious chemical rule; however, again we see that multipoles show a more complicated relationship to the underlying method.

Generally, CCSD errors can be substantial for multipoles in some instances, suggesting that perturbative triplets induce a large change. As previous authors \cite{Hait2018} have noted, we see this as a cause for concern, as it may indicate that correlation beyond triple excitations significantly affects multipoles.
At the CCSD level, the dipoles of 15 species exhibit errors between \SIrange{5}{10}{\%} and the quadrupoles of 7 species exhibit errors between \SIrange{5}{7}{\%}. Again, we could not identify simple chemical rules to associate these outliers.
These species are hence subject to subtle uncertainties.
Nevertheless, these possible residual errors in our CCSD(T) data will be much smaller than the relative errors from DFAs, which we analyze in the next section. 

Thus far, we have used the anisotropic error metric in comparisons. 
The isotropic errors are similar to the previously discussed anisotropic ones, as shown in Figure \ref{fig-si:error_distribution-iso}. The RMS anisotropic errors are virtually identical to isotropic ones for polarizabilities and about \SIrange{10}{25}{\%} larger for multipoles. Hence, isotropic errors appear to be a generally sufficient indicator of performance. Furthermore, the general theme that multipoles are more intricate with respect to the electronic structure method repeats.

We note that the only pair of properties that shows reasonable correlation in $e_\text{aniso}$ are \qqpols and \ddpols (see figure \ref{fig-si:error-correl_btw-props}). 
While dipoles and quadrupoles show a highly similar distribution in Figure \ref{fig:error_distribution}, their individual errors are uncorrelated.
This again underpins that multipoles have more complicated origins than polarizabilities. 

\subsubsection{Performance of DFA}
\begin{figure}
    \centering
    \includegraphics[width=\linewidth]{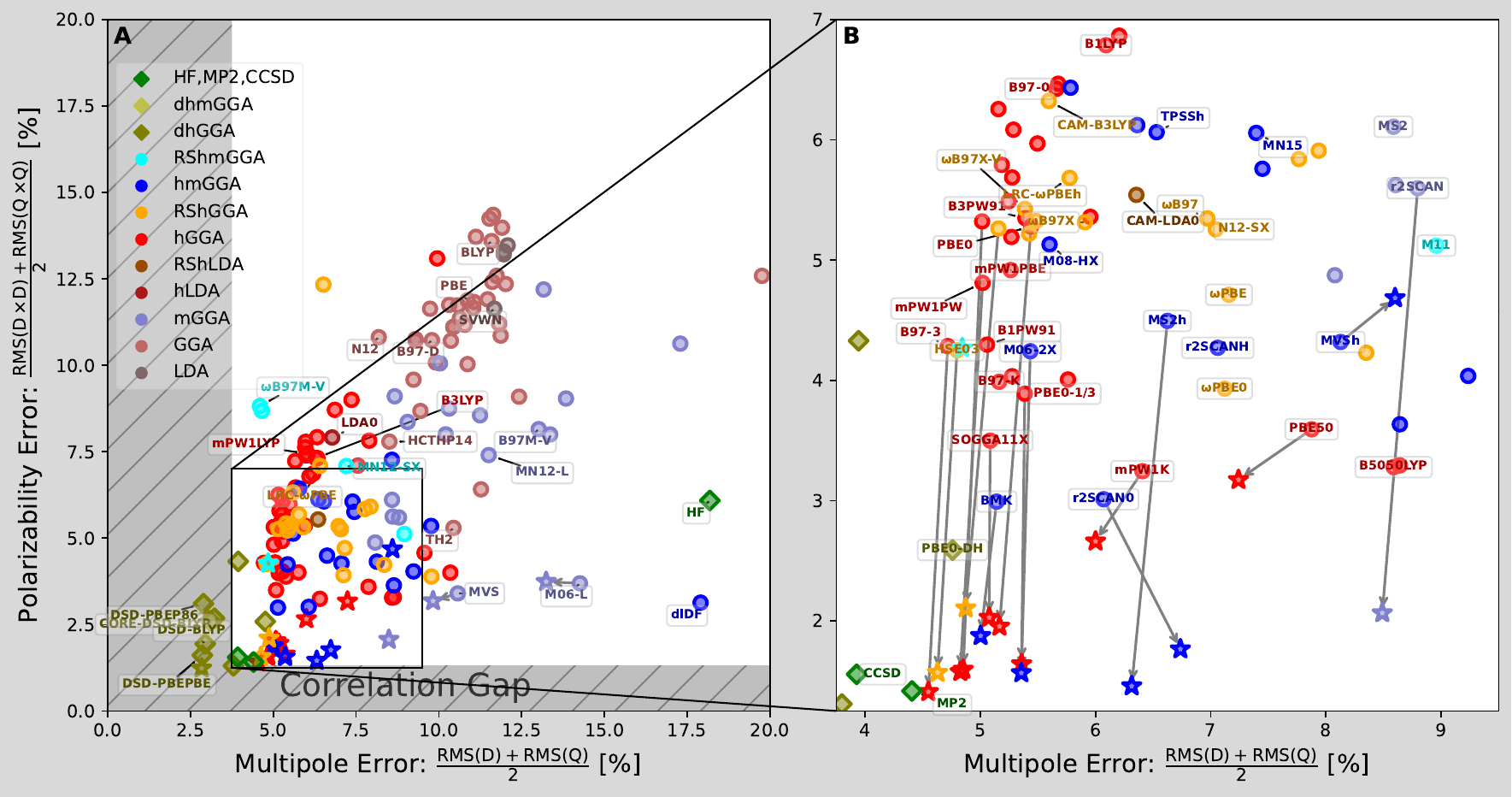}%
    \caption{Aggregate polarizability against aggregate multipole RMSE, with each dot a DFA colored according to its class. Stars mark the GRAC version of the corresponding base DFA from which the arrow originates.
    For labeled DFA, see Table \ref{tab:methods_leaderboard} for performance scores.
    \label{fig:method_performance}
    }
\end{figure}

Unlike the systematic behavior of molecular properties as a function of level of correlation, there is -- at least at first glance -- no similarly systematic behavior amongst the DFAs. 
As we will see, any given DFA might perform vastly differently across different properties.
While one can choose a DFA based on the property of interest, we aim to identify DFAs that perform consistently well across multipoles and polarizabilities at both dipolar and quadrupolar ranks. Such DFAs would provide a holistic description of intermolecular interactions and would not only be good candidates for databases involving higher-ranking terms, but also better suited to intermolecular energy calculations using SAPT(DFT).
Therefore, we perform a holistic analysis of DFA performance for both multipoles and polarizabilities at both dipolar and quadrupolar ranks.

Figure \ref{fig:method_performance} shows our performance indicator \rmsanis for multipoles and polarizabilities --- both aggregated by rank --- across a wide range of DFAs.
This correlation plot allows us to classify DFAs simultaneously by their performance on multipoles and polarizabilities.
We observe that DFAs can cluster by class (indicated by color), but performance does not strictly improve along Jacob's ladder.
Performance of individual DFAs can be found in Table \ref{tab:methods_leaderboard}.

As seen in Figure \ref{fig:method_performance}A, both rung 1, Local Density Approximation functionals (LDAs), and rung 2, the Generalized Gradient Approximation functionals (GGAs), are densely clustered with relatively large errors of about \prc{10} for both multipoles and polarizabilities.
At rung 3 of Jacob's ladder, meta-GGAs (mGGAs) show widely distributed errors, particularly for multipoles, where errors can exceed those of GGAs, but they substantially improve polarizabilities.
Rung 4 of Jacob's ladder introduces some fraction of exact exchange, leading to hybrid and range-separated DFA classes. 
The hybrid GGAs (hGGAs) and hybrid mGGAs (hmGGAs) generally improve over GGAs for both polarizabilities and multipoles.
The multipolar errors of hGGAs cluster tightly around the \prc{5} mark, while the polarizability errors are more spread out between \SIrange{2.5}{7.5}{\%}.
For hmGGAs, we also see large improvement over the mGGAs, but errors in multipoles can still be large and are spread out relatively widely between \SIrange{6}{11}{\%}. However, errors in the polarizabilities are smaller than those of hGGAs and cluster tightly between \SIrange{2.5}{5}{\%}. Moreover, multipoles of hGGA remain, on average, better than multipoles of hmGGAs.

Range separation has been shown to be essential for fixing self-interaction error in systems with notable charge delocalization and is expected to yield the correct long-range form of the exchange potential, which is crucial for accurate polarizabilities and higher-ranking multipoles.
However, on average, range-separated GGAs or hGGAs (collectively termed RShGGAs) show negligible impact on polarizabilities and even tend to worsen multipoles. The few range-separated mGGAs or hmGGAs (collectively termed RShmGGAs) considered tend to worsen performance, particularly for polarizabilities. For RShGGAs, average errors on multipoles lie between \SIrange{5}{8}{\%}, and errors in polarizabilities are clustered around \prc{4}.

In Figure \ref{fig:method_performance}A, we have highlighted what we call a ``correlation gap'' that bounds the errors made by DFAs up to and including rung 4 of Jacob's ladder. Specifically, all these DFAs show errors larger than \prc{4.5} for multipoles and larger than \prc{2} for polarizabilities, and cluster remarkably sharply before these boundaries.
These are about the performance margins of CCSD and MP2, which, in a sense, act as bounds for these DFAs.
Only double-hybrid DFAs, which constitute the fifth and last rank of Jacob's ladder, can penetrate this gap. They show excellent errors of about \prc{2.5} on multipoles and \prc{1.5} on polarizabilities. 
These DFAs introduce explicit correlation through MP2 and are much more computationally expensive.

A systematic way to improve DFA performance for intermolecular interactions is to use asymptotic corrections. As discussed in the theory section, asymptotic correction is based on a molecule-specific shift. In Figure \ref{fig:method_performance} we see that GRAC improves the performance of most DFAs, mostly for polarizabilities.
Indeed, DFAs such as B97-3, HSE03, and PBE0, which already perform well for multipoles, are improved to achieve near-on-par performance with MP2 and CCSD for polarizabilities.
However, with some exceptions, the asymptotic correction makes virtually no change to multipole errors.

We previously analyzed the relative magnitudes of \rmsanis between multipoles and polarizabilities by aggregating over rank; now we also want to comment on the same picture within multipoles and polarizabilities of different ranks.
Figure \ref{fig-si:dfa_performance_correlation} shows the \rmsanis of dipoles against quadrupoles and \qqpols against \ddpols. 
For multipoles, the range of errors made by various DFAs is similar for dipoles and quadrupoles, but the scatter is large. So a DFA that works well for one cannot be assumed to work well for the other. For example, BMK scores \rmsanis values of \prc{4.8} on dipoles and a slightly larger \prc{5.4} on quadrupoles, while B97-3-AC scores \prc{5.4} on dipoles and \prc{3.7} on quadrupoles.
Figure \ref{fig-si:dfa_performance_correlation}C and D, on the other hand, show that the \rmsanis of \qqpols and \ddpols are strongly correlated. Hence, DFA performance on \ddpols is a reasonable indicator of DFA performance on \qqpols. 
However, for non-asymptotically corrected DFAs, the magnitude of \rmsanis for \qqpols is about twice as large as for \ddpols. This suggests that polarizability errors of such DFAs might increase with multipolar rank, with larger errors possible for, e.g., octopole-octopole polarizabilities.
But with the asymptotic correction, this changes, and DFAs now show similarly small errors in both \ddpols and \qqpols. For example, for B97-3-AC we obtain errors of \prc{1.3} and \prc{1.5} for the two ranks of polarizabilities. s

MP2 and CCSD have proven quite consistent for properties and are relatively close to the CCSD(T) reference. DFAs, on the other hand, show highly varied performance, and while introducing exact exchange and the asymptotic correction can improve their performance dramatically, there is still a wide spread in error. Hence, we must choose our DFA carefully.

\subsubsection{Specific DFA recommendations}
We now highlight performance on specific DFAs, using their performance across all four properties as shown in Table \ref{tab:methods_leaderboard}.
Because of the correlation gap described above, we highlight errors relative to CCSD: green indicates proximity to CCSD, and red (tending to grey) indicates deviations.

No contenders appear within the local and semilocal DFAs. Only hybrid DFAs perform acceptably. Among these, only the asymptotically corrected DFAs perform well for all four properties. The best among these is the hybrid GGA B97-3-AC, which is closely followed by HSE03-AC.
These are followed by SOGGA11X-AC and PBE0-AC, the latter has extensive application for intermolecular interactions through SAPT(DFT).\cite{Hesselmann2004,Misquitta2005}
For extended conjugated systems, one requires long-range separated HF exchange. The best range-separated DFA, HSE03-AC, uses short-range-separated HF exchange,\cite{Henderson2008} which makes it unsuitable for these systems. Unfortunately, long-range-separated DFAs, including commonly used ones like Cam-B3LYP and $\omega$B97M-V, perform comparatively poorly relative to hGGA, particularly for quadrupoles and \qqpols.
Amongst the hybrid meta-GGAs, BMK is the best overall DFA, with the asymptotically corrected variant, BMK-AC, inducing only a slight improvement over the standard DFA. 
In contrast, the dispersionless density functional, dlDF, exhibits quite accurate polarizabilities, but its multipoles show the largest errors in our analysis, with \prc{13} on the dipoles and \prc{24} on the quadrupoles. These errors are comparable to those made by HF, which likely reflects the large fraction of exact exchange of \prc{61}.

We considered only three range-separated hmGGAs, $\omega$B97M-V, MN11, and MN12-SX, and one asymptotically corrected variant, $\omega$B97M-V-AC.
Of these, only $\omega$B97M-V performs reasonably on multipoles, but polarizabilities are poor even for the asymptotically corrected version, with errors substantially higher than those of B97-3-AC.

As noted above, double hybrids are the only class of DFAs that surpass CCSD in accuracy. Indeed, for all considered double hybrids, particularly DSD-PBEPBE, we see errors of around \prc{3} on the multipoles and \prc{2} on polarizabilities. 
However, the high accuracy of properties evaluated using such double hybrids comes at a computational cost that is equivalent to that of MP2.

\subsubsection{Systematic errors}

The distribution of $e_\text{aniso}$ of the DFA shown in Figure \ref{fig:error_distribution} (find an analogous figure for $e_\text{iso}$ in Figure \ref{fig-si:error_distribution-iso}) generally shows mostly balanced systematic errors for multipoles (panels A and B) as the \mseis lies close to zero, except for the dipoles of PBE (see panel A).
For polarizabilities, we see strong overestimation for PBE0, $\omega$B97M-V, and particularly PBE. 
Generally, GGAs appear to underestimate dipoles strongly and quadrupoles slightly, while polarizabilities are strongly overestimated, as can be seen in Table \ref{tab:methods_leaderboard} (see \ref{fig-si:rms_vs_systematic_errors} for indication of spread through all DFAs sampled). For hGGAs, \mseis is much more balanced, and as we will show in the following subsection, this might arise from systematic compensation between under- and overestimation between the HF-character (entering through exact exchange) and GGAs.

Figure \ref{fig-si:rms_vs_systematic_errors} shows that \rmsanis and \mseis are strongly correlated for polarizabilities, indicating that a large part of the observed \rmsanis arises from systematic errors. For dipoles, the correlation between \rmsanis and \mseis is much weaker and virtually non-existent for quadrupoles.
The correlation of these two error metrics suggests that scaling might be a route to error reduction, and we will explore this in Sec.~\ref{sec:scaling}. 

\subsubsection{Analysis of exact exchange for DFA}
\begin{figure}
    \centering
    \includegraphics[width=0.85\linewidth]{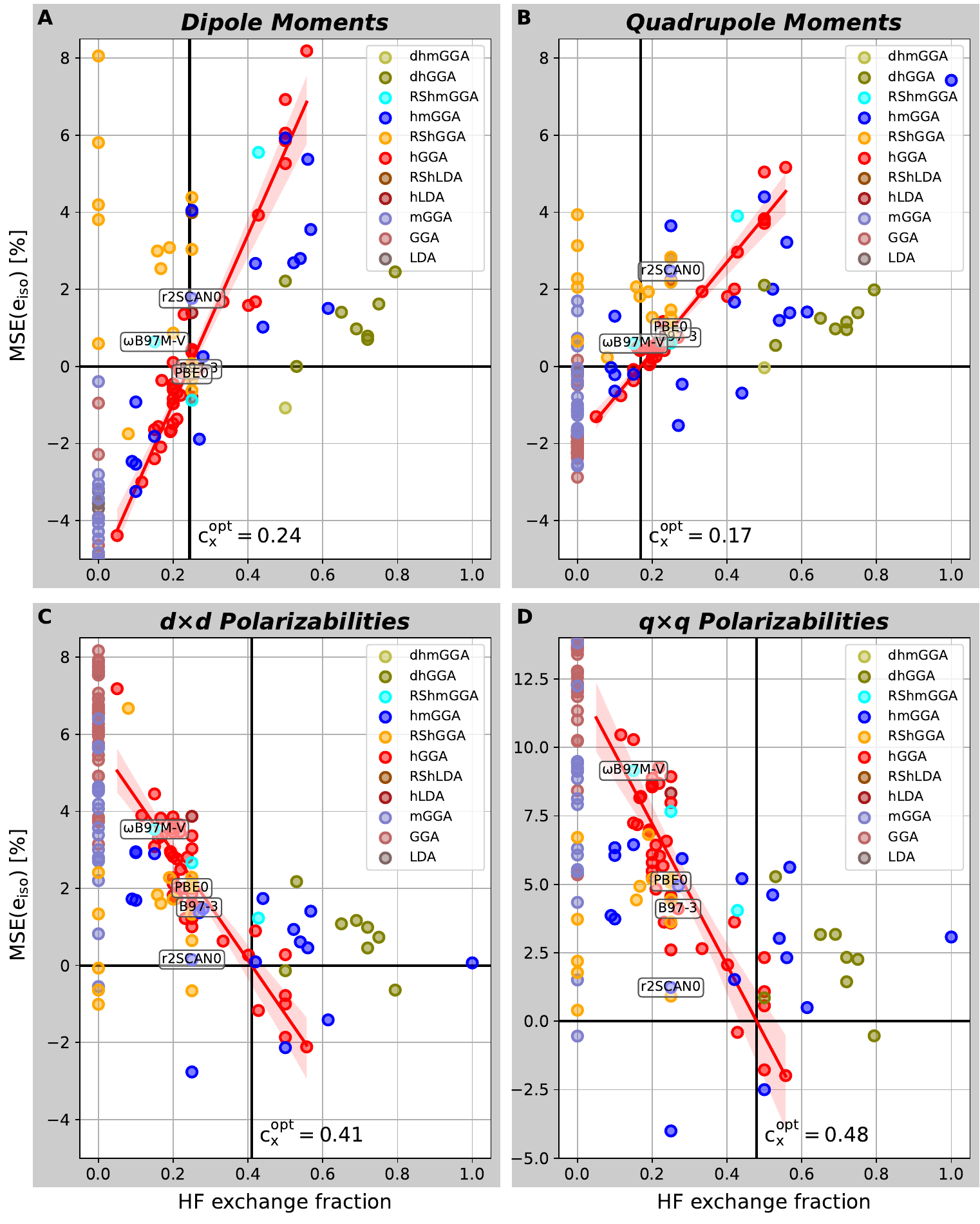}
    \caption{\mseis against the amount of exact short-range exchange \chf. 
    Linear fits for global hGGAs shaded with a \prc{95} confidence interval through the \texttt{regplot} function of the \texttt{python} library \texttt{seaborn}.
    \label{fig:properties_vs_exchange_avriso}
    }
\end{figure}

In the previous analysis of DFA classes according to Jacob's ladder, we saw that the addition of exact exchange had the most  
profound impact on properties. In the following, we investigate the role of \chf, the precise fraction of short-range HF exchange \hfsr, with respect to each property.

Figure \ref{fig:properties_vs_exchange_avriso} plots \mseis against \chf for the DFAs surveyed. 
We see that hGGAs follow a linear trend while the remaining DFA classes show scattering. 
Like GGAs ($\text{\chf}=0$), hGGAs with small \chf underestimate the magnitude of multipoles ($\text{\mseis}\lt 0$) and overestimate the magnitude of polarizabilities ($\text{\mseis}\gt 0$).
As \chf increases, \mseis crosses zero for all properties. The position of this zero-crossing -- which indicates the amount of \chf at which the modulus of the systematic error is minimized -- varies across properties:
$\text{\chf}=0.24$ for dipoles, $\text{\chf}=0.17$ for quadrupoles, $\text{\chf}\approx 0.4$ for \ddpols, and $\text{\chf}\approx0.5$ for \qqpols.

In contrast to the linear trend of \mseis with respect to \chf, the \rmsanis of hGGAs shown in Figure \ref{fig-si:properties_vs_exchange_rmsaniso},  follows a parabolic trend. 
This trend is strong for dipoles (panel A) and \ddpols (panel C), as indicated by the shaded confidence intervals, while for quadrupoles (panel B) and \qqpols (panel D) the trend is ambiguous. 
The minima lie close to the optimal \chf suggested previously: \SI{.3}{} for dipoles, \SI{.4}{} for \ddpols, while the best-performing hGGA for quadrupoles and \qqpols occur around \SI{0.2}{} and \SI{.5}{}, respectively.
For quadrupoles, one might instead explain the distribution of \rmsanis for hGGA around $0.1\lt\text{\chf}\lt 0.3$ as a bimodal clustering (as indicated by the two ellipses in Figure \ref{fig-si:properties_vs_exchange_rmsaniso}B): large errors around \prc{6.5} for the hGGAs and RShGGAs using LYP, as well as the hybrid LDA LDA0 and range-separated LDA Cam-LDA0, and smaller errors of \prc{4.5} for the remainder (see Figure \ref{fig-si:properties_vs_exchange_rmsaniso}B).

As we previously noted in our holistic performance analysis, many hGGAs tend to outperform hmGGAs, RShmGGAs, and RShGGAs. Our analysis suggests that the best-performing hGGAs for multipoles show $\text{\chf}\approx 0.25$, which is optimal for dipoles and near-optimal for quadrupoles. 
For polarizabilities, however, a larger fraction of exact exchange is needed, with an optimal value of $\text{\chf}\approx 0.5$. This can be explained by an increase of the HOMO--LUMO gap with \chf, which strongly affects polarizabilities; this can also be addressed with the asymptotic correction \cite{Tozer1998}.
Consequently, to {\em simultaneously} compute accurate multipoles and polarizabilities, we may use $\text{\chf} = 0.25$ {\em while} applying an asymptotic correction like GRAC \cite{Gruening2001}.
This is indeed borne out in our analysis, as B97-3-AC ($\text{\chf} = 0.27$) and PBE0-AC ($\text{\chf} = 0.25$) are some of the best-performing DFAs for all properties.

The previously noticed bimodal distribution of errors for the quadrupoles seen in Figure \ref{fig-si:properties_vs_exchange_rmsaniso}B arises almost entirely from the choice of correlation functional; the LYP correlation functional yields systematically larger errors than the PW91 correlation functional.
This is well illustrated by the drop of \rmsanis for quadrupoles from \prc{6.5} for B3LYP to \prc{3.9} for B3PW91 and likewise from \prc{6.8} for B1LYP to \prc{4.0} for B1PW91. These DFA pairs share an equivalent parametrization, differing only in the correlation potential \cite{Becke1993,Stephens1994,Adamo1997}.
By contrast, the difference between the B1 and B3 exchange kernels accounts for only a \prc{.5} difference in error. 
For other DFA classes, we see no evidence of strong systematic behavior with respect to the nature of the exchange-correlation potential.

\subsection{Basis sets for DFT}
\begin{figure}
    \centering
    \includegraphics[width=\linewidth]{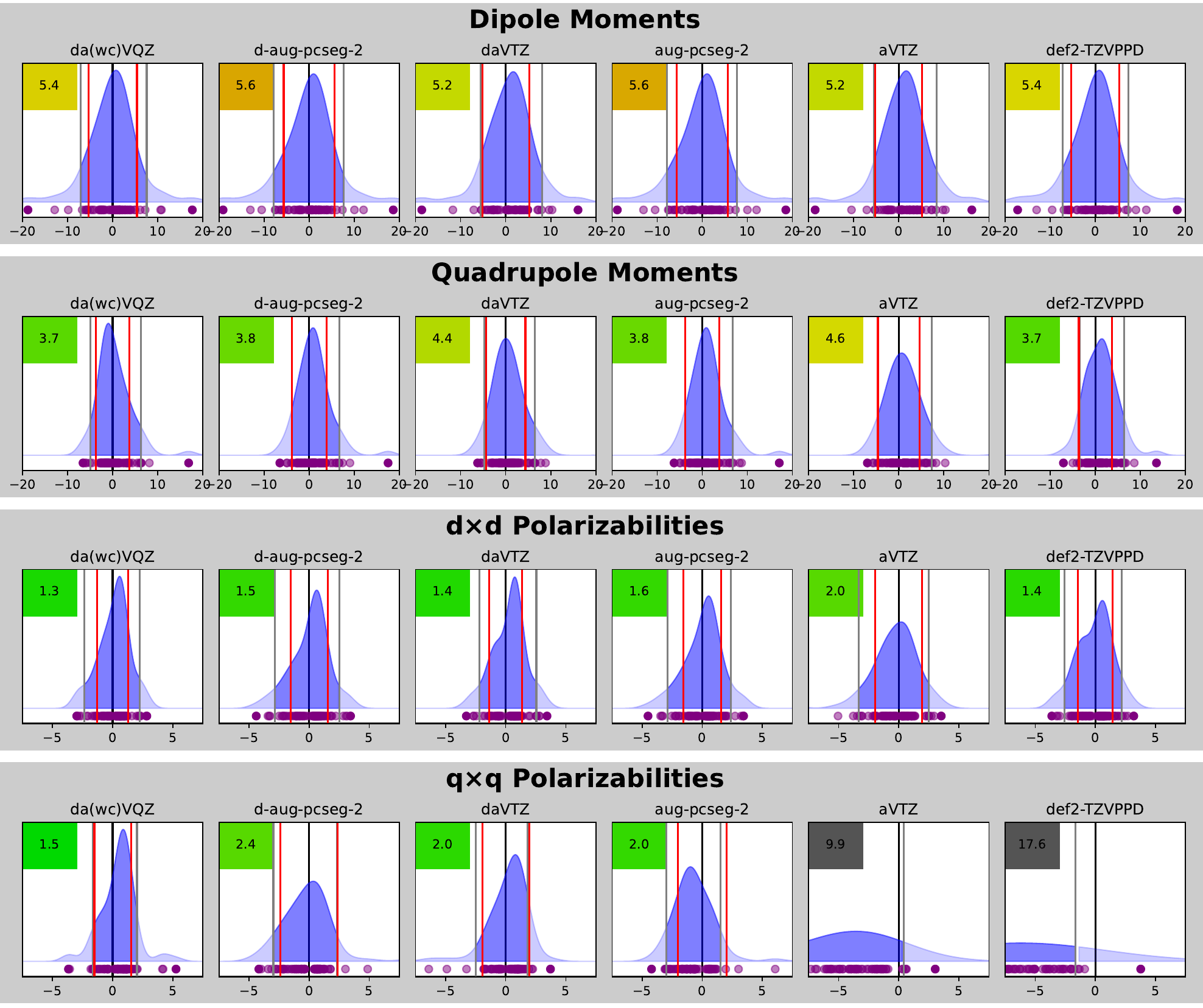}
    \caption{
    Distribution of signed anisotropic relative error at B97-3-AC with different basis sets --  all against CCSDT/\mainbasis -- for each property. 
    The dark blue shading in the distribution estimation indicates a \prc{90} confidence interval.
    The numbers in the top right-hand corner are the RMSE of the distributions (also the vertical red line within the distribution) and are color-coded according to the deviation, with green indicating the minimum value and red indicating \prc{8}.
    \label{fig:dft_basis_error}
    }
\end{figure}
\begin{figure}
    \centering
    \includegraphics[width=.75\linewidth]{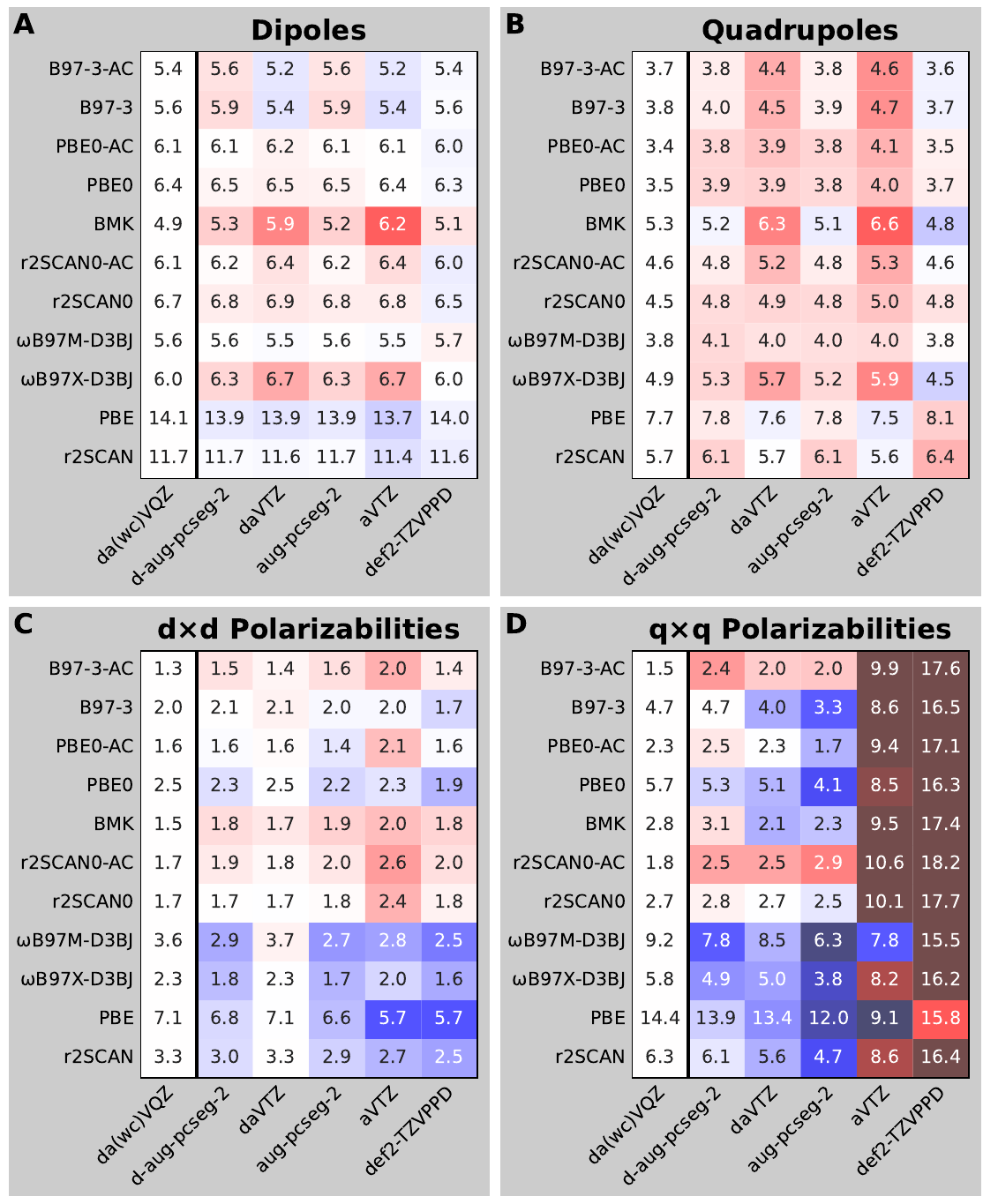}
    \caption{RMSE of properties evaluated at certain methods and basis sets against CCSD(T)/\mainbasis. The color scheme indicates the deviation from DFT/\mainbasis RMSE, with dark red at \prc{+3} to dark blue at \prc{-3}.   
    \label{fig:basis_error_table}
    }
\end{figure}

For many applications, \mainbasis (defined at the end of section \ref{sec:cc_basis_set_convergence}) is too computationally costly, motivating the use of smaller basis sets. 
We quantify the resulting inaccuracies by the change in $e_\text{aniso}$ relative to our CCSD(T)/\mainbasis references, evaluated over the same compounds used in the DFA performance analysis above. We consider a wider variety of {\em Dunning}, {\em Jensen}, and {\em second-generation Karlsruhe } basis sets; specific basis sets can be found in Figure \ref{fig:basis_error_table} and Figure \ref{fig-si:basis_error_table}.

Figure \ref{fig:dft_basis_error} shows how the distribution of $e_\text{aniso}$ changes across compounds for the example of B97-3-AC, using the following triple-$\zeta$ basis sets alongside \mainbasis: d-aug-pcseg-2, daVTZ, aug-pcseg-2, aVTZ, and def2-TZVPPD.
In most cases, both the distribution of $e_\text{aniso}$ and \rmsanis deviate only slightly from those obtained with \mainbasis, with the distribution peaks shifting only slightly and the \rmsanis changing slightly (the latter is highlighted by the color scheme). 
Dipoles, quadrupoles, and \ddpols show only small \rmsanis\ changes relative to \mainbasis: \prc{.2} for dipoles and \prc{.7} for quadrupoles and \ddpols. For \qqpols, however, aVTZ and def2-TZVPPD incur unacceptably large errors of \prc{10} and \prc{18}, respectively.

Consequently, only d-aug-pcseg-2, daVTZ, and aug-pcseg-2 yield acceptable errors across all four properties.
Among these, aug-pcseg-2 is the smallest (Table \ref{tab-si:count_basis_functions}) and is therefore the recommended choice. We note that def2-TZVPPD (and the smaller def2-TZVPD) performs at least as well as aug-pcseg-2 for every property except \qqpols, while using \prc{20} fewer basis functions for first- and second-row elements (Table \ref{tab-si:count_basis_functions}).

Figure \ref{fig:basis_error_table} extends the analysis of B97-3-AC to the hybrid DFAs B97-3-AC, B97-3, PBE0-AC, PBE0, r2SCAN0-AC, r2SCAN0, $\omega$B97M-D3BJ, $\omega$B97X-D3BJ, as well as the the semi-local DFAs r2SCAN-AC, r2SCAN and PBE. 
Red shading marks an increase in \rmsanis relative to \mainbasis, blue a decrease - the latter being beneficial, as it delivers improved accuracy at lower cost; we will discuss this error cancellation further below.
All triple-$\zeta$ basis sets shown produce generally small changes in \rmsanis for dipoles (panel A), quadrupoles (panel B), and \ddpols (panel C).
For \qqpols (panel D), aVTZ and def2-TZVPPD generally fail to achieve errors below \prc{9} and \prc{17}, respectively; the remaining basis sets occasionally increase \rmsanis\ slightly but far more often reduce it substantially.
Similarly, \rmsanis for \ddpols (panel C) are predominantly improved by smaller basis sets, though by smaller margins, whereas dipoles (panel A) and quadrupoles (panel B) predominantly worsen, particularly for daVTZ and aVTZ.  Similarly, an investigation of real-space density has shown that aug-pc-seg-$n$ basis sets outperform their corresponding aV$n$ counterparts.\cite{Gubler2025}
Overall, these trends resemble the previous specific analysis of B97-3-AC and reinforce the recommendation of aug-pcseg-2.

Figure \ref{fig-si:basis_error_table} additionally reports data for double- and quadruple-$\zeta$ basis sets for the DFAs discussed above. None of the double-$\zeta$ basis sets is acceptable for all four properties, though d-aug-pcseg-1, aug-pcseg-1, taVDZ, and daVDZ yield reasonable \rmsanis for dipoles and \ddpols.  
Quadruple-$\zeta$ basis sets perform very similarly to \mainbasis; however, only aug-pcseg-3 performs well on \qqpols, while def2-QZVPPD does not show \rmsanis under \prc{14}, and aVQZ shows both under- and overestimation of up to \prc{5}.

\subsubsection{Error cancellation and scaling}\label{sec:scaling}

As previously shown in Figure \ref{fig:basis_error_table}, smaller basis sets occasionally lower \rmsanis. A particularly striking example is the \qqpol from PBE/aVTZ, which yields \rmsanis of \prc{9.5}, compared to \prc{14} for \mainbasis. 
This behavior may result from cancellation between the systematic overestimation of \qqpols by PBE and the systematic underestimation from less diffuse basis sets. While we do not observe such error cancellation for dipoles and only occasionally for quadrupoles, it occurs frequently for polarizabilities, as shown in Figure \ref{fig-si:basis_error_table}. Note that the actual basis-set errors when comparing DFA at \mainbasis with smaller basis sets are generally much larger, as shown in Figure \ref{fig-si:cancellation_of_error}.

In general, a scaling factor $c$ can correct systematic errors in computed properties.
We define scaled properties $\theta^\text{DFT}_\text{scaled}$ with respect to the following $c$:
\begin{align}
    c&=\langle e_\text{iso}(\theta^\text{DFT}, \theta^\text{CC} ) \rangle
    \\ \theta^\text{DFT}_\text{scaled} &= c^{-1} \theta^\text{DFT}
\end{align}
By construction, \mseis of $\theta^\text{DFT}_\text{scaled}$ is zero. 

We apply this scaling to the methods previously shown in Figure \ref{fig:basis_error_table}, using the corresponding scaling factors listed in Table \ref{tab-si:scaling_factors}. The comparison of \rmsanis of scaled and unscaled properties is shown in Figure \ref{fig-si:basis_scaling_table}.
While scaling shows negligible improvement for multipoles, it can markedly improve polarizabilities.
This improvement is particularly pronounced for methods with high initial \rmsanis for polarizabilities, such as PBE or $\omega$B97M-V, where the error is halved. However, methods with asymptotic correction and r2SCAN0 see virtually no improvement.
We note that these improvements might be overly beneficial since the scaling factors were optimized for the same data set used for validation. Nevertheless, we expect the improvements through scaling to generalize reasonably well 
since they absorb inherent systematic errors in both methods and the basis set.

\section{
    Conclusions
}\label{sec:conclusions}This study introduces novel references for a holistic assessment of electronic-structure methods using molecular dipole and quadrupoles, as well as dipole-dipole and quadrupole-quadrupole polarizabilities (\ddpols and \qqpols) at the CCSD(T) level for 73 non-spin-polarized small molecules.
These properties are important, inter alia, for higher-order electrostatics and dispersion models that include contributions beyond the usual leading-order terms.
We analyze basis-set requirements, the impact of correlation, and the performance of 176 density functional approximations (DFAs), with special focus on the role of the asymptotic correction and choices of correlation DFAs.
We also propose a framework for choosing finite-field values, which is essential for accurate polarizabilities through finite-field calculations.

The basis set requirements for both correlated methods and DFAs are governed by the highest-rank properties.
\Qqpols and quadrupoles are more demanding on basis sets than \ddpols and dipoles, with \qqpols strictly requiring doubly-augmented basis sets.
Core polarization plays a minor role for properties of molecules containing first- and second-row atoms, but the quadrupoles of molecules containing third-row species show high sensitivity to core polarization.
For benchmark generation with correlated methods, to strike a balance between accuracy and computational cost, we recommend a composite basis, which we refer to as the da(wC)VQZ basis. This basis uses the d-aug-cc-VQZ basis sets for first- and second-row atoms, and 
d-aug-cc-pwCVQZ for third-row atoms (possibly with the exception of chlorine).
For properties computed with DFAs, the {\em Jensen} basis set aug-pcseg-2 strikes a good balance as it introduces minimal error relative to the much larger basis sets we have used, and it should be preferred over aug-cc-pVTZ or def2-TZVPPD bases.

In terms of method accuracy: 
CCSD and MP2 perform similarly across all properties, with errors of about \prc{4} on multipoles and \prc{1.5} on polarizabilities. Double-hybrid DFAs surpass this performance for multipoles, particularly DSD-PBEPBE.
All other DFA classes are bounded in accuracy by CCSD and MP2. This gap, which we term a ``correlation gap'', is widest for the multipoles, where even the best DFAs show errors of more than \prc{4.5} against our reference. For polarizabilities, this gap is smaller, about \prc{2}, and drops to about \prc{1} through application of an asymptotic correction.  We note that the asymptotic correction does not appear to reduce multipole errors significantly. 

For each property, a DFA can be found that closely approaches CCSD in performance, but in keeping with the holistic nature of our assessments, we rank DFAs by their ability to reproduce all four quantities together. We motivated this requirement in the Introduction.
If one excludes double hybrids because of their higher computational cost, the best DFA is the asymptotically corrected hybrid GGA B97-3-AC, which performs largely on par with CCSD. 
This DFA appears to slightly outperform PBE0-AC, which has seen long use in SAPT(DFT) \cite{Misquitta2005,Hesselmann2004}.
The best method without asymptotic correction is the hybrid meta-GGA, BMK, although it results in somewhat larger quadrupole errors (\prc{5.4}).
Surprisingly, none of the tested DFAs with long-range-separated exchange performs well across all properties. 
With the best being the range-separated hybrid meta-GGA $\omega$B97M-V, it yields multipoles with errors just slightly larger than those from B97-3-AC, but it gives large errors in the \ddpols and \qqpols (\prc{3.8} and \prc{9.3}). 
We note that these errors might be reduced by scaling the polarizabilities.
Therefore, we recommend B97-3-AC with the aug-pcseg-2 basis over any of the DFAs in use today.

We find exchange to be important for properties and have searched for correlations between property accuracy and the fraction of short-range exchange.
For multipoles, $17-24$\% HF exchange is optimal, and this agrees with the fraction of HF exchange in commonly used hybrids. 
For polarizabilities, the optimal values are larger, between $41$\% and $48$\%, consistent with the dependence of polarizabilities on the excitation spectrum in addition to the density tails. 
We suggest that the asymptotic correction provides a convenient way to satisfy the optimal HF exchange condition for the multipoles while simultaneously correcting the excited states to yield accurate polarizabilities.

These observations underpin preceding observations in the literature\cite{Sim2022, Medvedev2017, Brorsen2017,Hait2018}: modern DFAs do not generally provide an accurate description of the density and its response.
We therefore suggest that DFAs that perform well on intermolecular benchmarks, like $\omega$B97M-V, do not provide an accurate physical framework, which would hinder mapping onto accurate force-field-like terms.

Finally, we note that we studied only standard DFAs and those with an asymptotic correction. We have shown that standard range-separated DFAs do not perform well on all properties simultaneously, but we have not investigated tuned range-separated DFAs.
Range-separation with tuning is, in a sense, an analog of the asymptotic correction, and molecule-specific tuning of the range-separation parameter could improve this otherwise disappointing DFA class. 
This possibility is appealing as range-separation is known to be needed to control the self-interaction error within the monomer, particularly for large, heavily delocalized systems. 

\section{Acknowledgements} 
We acknowledge UKRI grant EP/X036863/1and support from the   PHYMOL project, funded from HORIZON-MSCA-2021-DN-01-01, Grant Agreement No. 101073474, as well as the research activities of the COST Action COSY (CA21101).
The authors acknowledge the support through computational resources by the GridPP Collaboration at Queen Mary University of London.

\section{Associated Content}
The computed properties are publicly available under: \url{https://github.com/bruno-von-bruening/higher-ranking_property_data}.

The Supporting Information includes: lists of the sampled compounds and DFAs approximations used; the finite-field methodology, including its setup and equations, along with validation against linear-response calculations; an assessment of errors arising from density fitting, together with remarks on the frozen-core approximation; and the performance of methods after scaling was applied. It also provides additional data visualizations that extend or complement those presented in the main manuscript.

\bibliography{Pol_Benchmarking}

\end{document}